\documentclass[preprint]{aastex61}

\newcommand{\kms}{km~s$^{-1}$}
\newcommand{\msol}{$M_{\sun}$}
\newcommand{\htwo}{H$_2$}
\newcommand{\feii}{[\ion{Fe}{2}]}
\newcommand{\core}{G53.11\_MM1}
\newcommand{\lhtot}{$L_{\rm H_{2}}$}

\newcommand{\lbol}{$L_{\rm bol}$}
\newcommand{\wm}{${\rm W\,m}^{-2}$}
\newcommand{\no}{\raisebox{.5ex}{\footnotesize\bf\#}} 
\newcommand{\nosm}{\raisebox{.5ex}{\tiny\bf\#}} 

\begin{document}

\title{A Parsec-scale Bipolar H$_2$ Outflow in the Massive Star Forming
Infrared Dark Cloud Core MSXDC G053.11+00.05 MM1\footnote{[RJS2006] MSXDC G053.11+00.05 MM1
or AGAL G053.141+00.069 in the SIMBAD database, operated at CDS, Strasbourg, France \citep{wenger00}}
\footnote{Based in part on data collected at Subaru Telescope, which is operated by 
the National Astronomical Observatory of Japan.}}

\correspondingauthor{Hyun-Jeong Kim}
\email{hjkim@astro.snu.ac.kr, koo@astro.snu.ac.kr}

\author[0000-0001-9263-3275]{Hyun-Jeong Kim}
\affiliation{Department of Physics and Astronomy, 
Seoul National University, \\
1 Gwanak-ro, Gwanak-gu, Seoul 08826, Republic of Korea}

\author{Bon-Chul Koo}
\affiliation{Department of Physics and Astronomy, 
Seoul National University, \\
1 Gwanak-ro, Gwanak-gu, Seoul 08826, Republic of Korea}

\author{Tae-Soo Pyo}
\affiliation{Subaru Telescope, National Astronomical Observatory of Japan, 
National Institutes of Natural Sciences (NINS), \\
650 North A`oh\=ok\=u Place, Hilo, HI 96720, USA}
\affiliation{School of Mathematical and Physical Science, 
The Graduate University for Advanced Studies (SOKENDAI), \\
Hayama, Kanagawa 240-0193, Japan}

\author{Christopher J. Davis}
\affiliation{Astrophysics Research Institute, Liverpool John Moores University, \\
146 Brownlow Hill, Liverpool, L3 5RF, United Kingdom}
\affiliation{National Science Foundation, 
2415 Eisenhower Avenue,
Alexandria, VA 22314, USA}

\begin{abstract}

We present a parsec-scale molecular hydrogen (H$_2$ 1-0 S(1) at 2.12~\micron) outflow
discovered from the UKIRT Widefield Infrared Survey for H$_2$. The outflow is located
in the infrared dark cloud core MSXDC G053.11+00.05 MM1 at 1.7~kpc
and likely associated with two young stellar objects (YSOs) at the center.
The overall morphology of the outflow is bipolar along the NE-SW direction with a brighter lobe
to the southwest, but the detailed structure consists of several flows and knots.
With the total length of $\sim$1~pc, the outflow luminosity is fairly high
with $L_{\rm H_{2}} > 6~L_{\sun}$, implying a massive outflow-driving YSO
if the entire outflow is driven by a single source.
The two putative driving sources, located at the outflow center, show
photometric variability of $\gtrsim$1~mag in {\it H}- and {\it K}-bands.
This, with their early evolutionary stage from spectral energy distribution (SED) fitting,
indicates that both are capable of ejecting outflows and may be eruptive variable YSOs.
The YSO masses inferred from SED fitting are $\sim$10~$M_{\sun}$ and $\sim$5~$M_{\sun}$,
suggesting the association of the outflow with massive YSOs.
The geometrical morphology of the outflow is
well explained by the lower mass YSO by assuming a single source origin, but without
kinematic information, the contribution from the higher mass YSO cannot be ruled out.
Considering star formation process by fragmentation of a high-mass core into
several lower mass stars, we also suggest the possible presence of another, yet-undetected
driving source deeply embedded in the core.

\end{abstract}

\keywords{ISM: individual objects ([RJS2006] MSXDC G053.11+00.05 MM1) --- 
ISM: jets and outflows --- stars: formation --- stars: protostars}

\section{Introduction}\label{sec:intro}

Outflows and jets from protostars are major outcomes of the star formation process 
and one of the prominent observational signs in star-forming regions. 
In low-mass star formation, outflows and jets, driven by magnetic stresses or 
magneto-centrifugal force in accretion disks, play an important role in 
removing a large fraction of angular momentum from rotating disks and 
provide a clue to accretion processes/history of young stellar objects (YSOs)
\citep[e.g.,][and references therein]{shu94,frank14,caratti15}.
In high-mass star formation, on the other hand, it is still controversial whether 
their formation process is a scaled-up version of low-mass 
star formation \citep{bonnell01,mckee03,wang10,tan14}, and 
the roles of outflows and jets have thus far remained unclear.
Since massive stars are small in number, distant (several kpc), heavily 
obscured ($A_{\rm V}$ up to 100~mag), and evolve in a short timescale 
compared to low-mass stars,
it is difficult to observationally examine massive star formation process.
Because of large extinction, outflows from massive YSOs are not 
accessible by optical emission lines (e.g., [\ion{O}{1}], [\ion{S}{2}], 
H$\alpha$), the outflow-shock tracers frequently used in low-mass YSOs, 
so they have mainly been  explored by molecular lines such as CO or SiO 
at (sub)millimeter wavelengths \citep[e.g.,][]{beuther02,wu05,lopez09}.
Those lines from radio observations trace molecular outflows 
but generally suffer from low spatial resolution except 
a few interferometer observations.
Recently, several surveys of outflows/jets in near-infrared (near-IR), 
particularly by using the \htwo\ 1-0\,S(1) line at 2.12~\micron\
have been carried out, allowing us to trace shocks in molecular 
outflows and investigate the primary outflows ejected from their driving sources 
on scales of a few thousands of AUs to parsecs.
Many studies have revealed \htwo\ outflows from intermediate- or high-mass YSOs 
some of which are well collimated as outflows from low-mass YSOs, 
suggesting that disk accretion is likely the leading mechanism 
in high-mass star formation as well as in low-mass star formation
\citep[e.g.,][]{davis08,davis10,varricatt10,lee13,caratti15}.

\begin{figure*}[t!]
\includegraphics[width=1\textwidth]{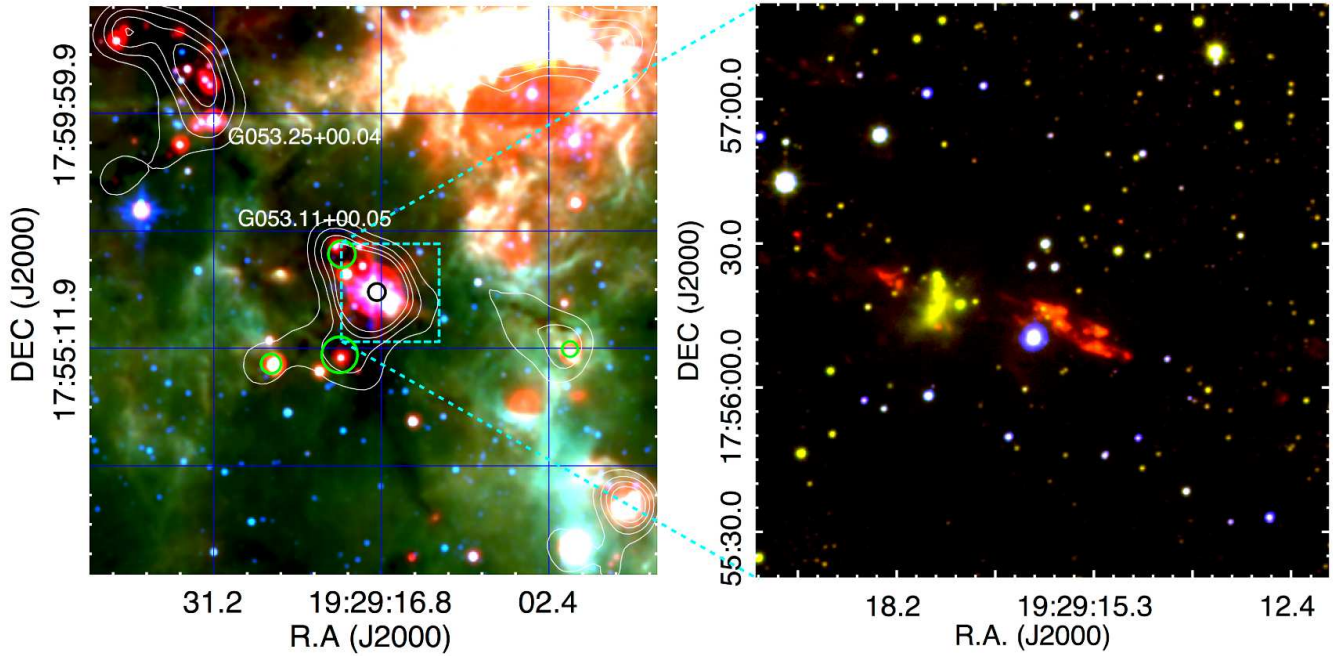}
\caption{Left: Three-color image of MSXDC G053.11+00.05 produced
from {\it Spitzer} IRAC 5.8 $\micron$ (B), IRAC 8.0 $\micron$ (G), and 
MIPS 24 $\micron$ (R) images.
White contours are the 1.2~mm continuum emission from the Bolocam 
Galactic Plane Survey \citep{bolocam}.
\core\ and the other four cores in MSXDC 
G053.11+00.05 are marked by black and green circles, respectively.
The north-east region is a part of another IRDC 
MSXDC G053.25+00.04 \citep{simon06}.
Right: Three-color image of the \core\ outflow marked by a cyan dashed-box 
in the left panel, produced from the UKIRT/WFCAM J(B), K(G), and \htwo(R) images. 
\label{fig1}}
\end{figure*}

In this paper, we present a remarkable \htwo\ outflow and 
putative outflow-driving YSOs discovered in the infrared dark cloud (IRDC) core 
MSXDC G053.11+00.05 MM1 \citep[\core\ hereafter;][]{simon06,rath06}
displayed in Figure~\ref{fig1}.
MSXDC G053.11+00.05 is a part of a long, filamentary CO molecular cloud 
located at Galactic coordinates $(l,b)\sim(53\fdg2, 0\fdg0)$, which 
was defined as IRDC G53.2 in our previous study 
\citep[see Figure~1 of][]{kim15}\footnote{\citet{ragan14} also identified 
the same molecular cloud GMF~54.0--52.0 but in a larger size.}.
The kinematic distance of IRDC G53.2 obtained from the CO line 
velocity of $\sim$23~\kms\ is from 1.7 to 2.0~kpc depending on the Galactic
rotation model \citep{rath06,ragan14,kim15}; in this study,
we adopt 1.7~kpc derived by using a flat rotation curve 
with $R_{\sun}=8.5$~kpc and $\Theta_{\sun}=220$~\kms\ \citep{kim15}.
IRDC G53.2 is an active star-forming region with more than 
300 YSO candidates \citep{kim15} and a large number of 
molecular hydrogen (\htwo\,1-0\,S(1) at 2.12~\micron) 
emission-line objects (MHOs) revealed from the UKIRT
Widefield Infrared Survey for \htwo\ \citep[UWISH2;][]{froebrich11,froebrich15}.
Among those MHOs identified in IRDC G53.2, 
the \core\ outflow we address here is the most prominent \htwo\ outflow 
with a well-defined bipolar morphology (Figure~\ref{fig1})
and is rather isolated from the central, crowded region where it is difficult
to speculate the driving source.
The \core\ outflow is likely associated with high-mass star formation 
as the outflow is found at the center of the IRDC core.

In MSXDC G053.11+00.05, five millimeter cores have been detected \citep{rath06}
as marked in the left panel of Figure~\ref{fig1}. Among them, \core\ is the brightest 
and most massive one with its mass of 124~\msol\ derived 
from the 1.2~mm flux \citep{rath06}.
At the center of the core, the bipolar \htwo\ outflow oriented in the NE-SW 
direction is located with two early-class (Class I) YSOs separated 
by $\sim$8$\arcsec$ \citep{kim15}, YSOs that are referred as 
YSO1 and YSO2 in this study.
Besides YSO1 and YSO2, there are about 80 mid-IR sources idetified 
in the Galactic Legacy Infrared Midplane Survey Extraordinaire (GLIMPSE) 
Catalog/Archive \citep{benjamin03,churchwell09} around the \htwo\ outflow, 
among which 19 sources are detected in all the {\it Spitzer} IRAC bands but 
not detected in the {\it Spitzer} MIPS 24~\micron\ band up to 8.4~mag except
one source included in \citet{kim15}.
The spectral indices calculated between 2 and 8~\micron\ \citep{lada87,greene94} 
mostly classify them as flat spectrum or Class II with a few Class I (Figure~\ref{fig2});  
the mid-IR colors \citep{gutermuth09} mostly classify them as photospheric sources 
or Class II. 
Although these Class I and II YSOs can drive the outflow,
the possibility that they are driving the \core\ outflow is low
because these YSOs are relatively far from the outflow center. 
Since the outflow is well defined by a bipolar shape,
the driving source is likely at the center of the outflow. Therefore, 
considering the central location as well as the early evolutionary class, we 
regard YSO1 and YSO2 as the putative driving sources of the \core\ outflow.

Toward \core, several maser detections have been previously reported:
22~GHz water maser, 44 and 95~GHz class I methanol masers 
from the Korean VLBI Network (KVN) observations \citep{kang15}; 
6.7 GHz class II methanol maser from 
the MERLIN observations \citep[G53.14+0.07;][]{pandian11}.
The detected masers with no radio continuum emission
at 5 GHz \citep{urquhart09} support star formation activity 
in early stages; the positional coincidence between
the 6.7 GHz methanol maser G53.14+0.07 at 
($\alpha_{2000}$, $\delta_{2000}$)=
(${\rm 19^{h}29^{m}17\,\fs581, +17\arcdeg56\arcmin23\farcs21}$) and
one of the two central YSOs (YSO1; see Section~\ref{sec:yso}) strongly
indicates that this YSO is a high-mass protostellar object. 
This suggests that either one (or both) of the central YSOs is massive and
a possible driving source of the outflow.

In this study, we investigate the characteristics of the \core\ outflow
and central YSOs using narrow- and broad-band IR imaging observational data.
We derive their physical parameters and discuss their properties. 
In Section~\ref{sec:data}, we present the observational data used in this study 
and data reduction process. In Section~\ref{sec:outflow}, we present the characteristics 
of the \htwo\ outflow by deriving the geometrical/physical parameters, 
and in Section~\ref{sec:feii}, 
we search for \feii\ emission associated with the \htwo\ outflow.
We then move to the central YSOs in Section~\ref{sec:yso} presenting 
their photometric variability and spectral energy distribution (SED) analysis. 
In Section~\ref{sec:origin}, we discuss the origin of the \core\ outflow  
based on the results from the foregoing sections, and we finally summarize and 
conclude our study in Section~\ref{sec:summary}.

\section{Data}\label{sec:data}

\subsection{UKIRT/WFCAM Wide-field Images}\label{sec:ukirt}

The outflow in \core\ was first identified from the UWISH2 survey. 
The UWISH2 survey mapped the First Galactic Quadrant 
($6\arcdeg \lesssim l \lesssim 65\arcdeg; |b| \lesssim 1\fdg5$) 
with the narrow-band filter centered on the \htwo\ emission line 
at 2.12~\micron\ using the Wide-Field Camera (WFCAM) at United Kingdom 
Infrared Telescope (UKIRT) from 2009 July to 2011 August.
The WFCAM has four Rockwell Hawaii-II HgCdTe arrays of $2048 \times 2048$ pixels 
and provides $13\farcm65 \times 13\farcm65$ field-of-view (FOV) images with 
a pixel scale of $0\farcs4$. The images are resampled to $0\farcs2$ 
in the final stacked images \citep{froebrich11}.
The IRDC G53.2 region was observed in 2010 and 2011.
For continuum subtraction from the narrow-band \htwo\ images,
we used the broad-band {\it K}-band images obtained in 2006 
from the UKIRT Infrared Deep Sky Survey of the Galactic 
plane \citep[UKIDSS GPS;][]{lucas08}.  

We also used the \feii\ images obtained from the UKIRT Widefield Infrared 
Survey for Fe$^+$ \citep[UWIFE;][]{lee14} to search for \feii\ emission 
associated with the \htwo\ outflow. UWIFE was designed to complement 
UWISH2 so that it covers the same area with the same instrument as UWISH2 
but using the \feii\ 1.644~\micron\ narrow filter. 
The UWIFE survey was performed through 2012 and 2013, and 
the \feii\ images of the IRDC G53.2 region were taken in 2012. 
During the observations, we obtained the {\it H}-band images as well 
for continuum subtraction considering possible variations of continuum emission
between 2006 (from UKIDSS GPS) and 2012. Details on the UWISH2 and 
UWIFE surveys are presented in \citet{froebrich11} and \citet{lee14}, respectively.

All WFCAM data were reduced by 
the Cambridge Astronomical Survey Unit (CASU) as described 
in detail in \citet{dye06}; 
astrometric and photometric calibrations \citep{hodgkin09} were carried out 
by using the Two Micron All Sky Survey (2MASS) catalogue \citep{2mass}.
Continuum subtraction from \htwo\ and \feii\ narrow-band images was conducted 
by using {\it H}- and {\it K}-band images, respectively, as follows. 
We first re-projected the broad-band image onto the corresponding narrow-band 
image to align their astrometry. Since the broad- and narrow-band filters have 
different bandwidths, we also scaled the broad-band image 
to match the flux of the narrow-band image.
Then, we performed point-spread-function (PSF) photometry of each image 
and removed detected point sources; we finally subtracted 
the point-source-removed broad-band image from the point-source-removed
narrow-band image to remove other extended continuum sources. 
The above method was developed as a part of the UWIFE data reduction 
process, and more detailed explanations are given in \citet{lee14}.

\subsection{Subaru/IRCS High-resolution Imaging Observations}\label{sec:subaru}

We performed near-IR imaging observations of the central part of the \core\ outflow 
with high angular resolution to explore the detailed structures of the outflow and 
the vicinity of the central YSOs. The observations were conducted on 2012 July 30 UT 
by using the Infrared Camera and Spectrograph \citep[IRCS;][]{tokunaga98,kobayashi00}
on the Subaru telescope in a service mode (ID: S12A0139S; PI: Pyo, T.-S.).
Combined with the adaptive optics (AO) system \citep[AO188;][]{hayano10},
IRCS provides near-IR (1--5~\micron) images with pixel scales of 20 and 52 mas per pixel 
for the FOV of $21\arcsec \times 21\arcsec$ and $54\arcsec \times 54\arcsec$, respectively. 
We obtained \feii\ 1.644~\micron, \htwo\ 2.122~\micron, 
{\it H} (centered at 1.63~\micron), and {\it K} ($K^{\prime}$ centered at 2.12~\micron)
images toward \core\ centered 
at ($\alpha_{2000},\delta_{2000}$)=(${\rm 19^{h}29^{m}17\,\fs29, +17\arcdeg56\arcmin17\farcs59}$) 
with a pixel scale of $0\farcs052$ (52 mas mode).
Total integration times were 4,500 s for narrow-band filters and 300 s for broad-band filters.
The AO guide star was at 
($\alpha_{2000},\delta_{2000}$)=(${\rm 19^{h}29^{m}16\,\fs178, +17\arcdeg56\arcmin10\farcs14}$),
about 18$\arcsec$ apart from the center of the observed field,
and the seeing after AO correction is $0\farcs17$ at {\it K}-band.
We reduced the IRCS data with IRAF\footnote{IRAF \citep{tody86,tody93} is distributed by 
the National Optical Astronomy Observatories, which are operated by 
the Association of Universities for Research in Astronomy, Inc., under cooperative 
agreement with the National Science Foundation.} 
and IRCS IRAF script package (ircs\_imgred) distributed by National Astronomical 
Observatory of Japan (NAOJ)\footnote{\href{http://www.naoj.org/Observing/DataReduction/index.html}
{http://www.naoj.org/Observing/DataReduction/index.html}} 
following the standard procedure including dark subtraction, flat-fielding, 
median-sky subtraction, dithered image alignment, and image combining. 
Continuum emission was subtracted from the narrow-band images (\feii\ and \htwo)
by using the broad-band images ({\it H} and {\it K}) with the same method 
applied for the UKIRT/WFCAM data. 

\subsection{Gemini/NIRI High-resolution Imaging Observation}\label{sec:gemini}

We also performed high-resolution {\it K}-band imaging observation of 
the central part of the \core\ outflow using the Near Infrared Imager and 
Spectrometer \citep[NIRI;][]{hodapp03} attached on the Gemini North 
telescope on 2015 August 29 UT (Program ID: GN-2015B-Q-16; PI: Lee, J.-J.).
Among NIRI's three cameras, we used f/32 camera with 
the Gemini facility AO system ALTAIR \citep{christou10}, which provides 
a pixel scale of $0\farcs022$ per pixel and a FOV of $22\arcsec \times 22\arcsec$.
We obtained {\it K}-band (Kshort filter centered at 2.15\micron)
images of the central region of the core centered at
($\alpha_{2000},\delta_{2000}$)=(${\rm 19^{h}29^{m}17\,\fs36, +17\arcdeg56\arcmin18\farcs32}$)
and the sky region, where there is no star, for background subtraction
with a total integration time of 720 s for each.
The AO guide star was the same as the one used in the Subaru/IRCS observations,
and the AO-corrected seeing is $0\farcs12$.
Data reduction was done with Gemini IRAF package and the python scripts 
for cleaning and linearity correction provided from Gemini 
Observatory\footnote{\href{http://www.gemini.edu/sciops/instruments/niri/data-format-and-reduction}
{http://www.gemini.edu/sciops/instruments/niri/data-format-and-reduction}}, 
by following the same standard procedure described 
in Section~\ref{sec:subaru}.

\subsection{Infrared Archival Data}\label{sec:archive}

Since \core\ has been identified as a point or compact source from near-IR to millimeter, 
we used mid- and far-IR archival data as complements to investigate the central YSOs. 
In mid-IR, we used {\it Spitzer} IRAC band (3.6, 4.5, 5.8, and 8.0~\micron) images from 
GLIMPSE\footnote{\href{http://www.astro.wisc.edu/glimpse/glimpsedata.html}
{http://www.astro.wisc.edu/glimpse/glimpsedata.html}} \citep{benjamin03,churchwell09} 
with the GLIMPSE I v2.0 Catalog/Archive, {\it Spitzer} MIPS 24~\micron\ image from 
MIPS GALactic plane survey (MIPSGAL; \citealt{carey09}), and
Wide-field Infrared Survey Explore \citep[WISE;][]{wright10} 
all-sky data\footnote{\href{http://wise2.ipac.caltech.edu/docs/release/allsky}
{http://wise2.ipac.caltech.edu/docs/release/allsky}}.
In {\it Spitzer} images, two YSOs separated by $\sim$8$\arcsec$ \citep{kim15}
are resolved but saturated in the MIPS 24~\micron\ image; 
in the WISE images, two YSOs are not resolved because of low angular resolution.
In far-IR, we used the {\it Herschel}\footnote{{\it Herschel} is an 
ESA space observatory with science instruments provided by European-led 
Principal Investigator consortia and with important participation from NASA.} 
Infrared Galactic Plane Survey \citep[Hi-GAL;][]{molinari10} data
and the catalog of the IRDC-associated starless and 
protostellar clumps with known distance in the Galactic longitude 
range $15\arcdeg \leq l \leq 55\arcdeg$ from Hi-GAL \citep{traficante15}
to extract the PACS 70~\micron\ flux of \core.

\section{Characteristics of the \htwo\ Outflow}\label{sec:outflow}

\subsection{\htwo\ Outflow Morphology}\label{sec:morphol}
\subsubsection{Identification of \htwo\ Emission}\label{sec:ident}

\begin{figure*}[t!]
\begin{center}
\includegraphics[width=0.9\textwidth]{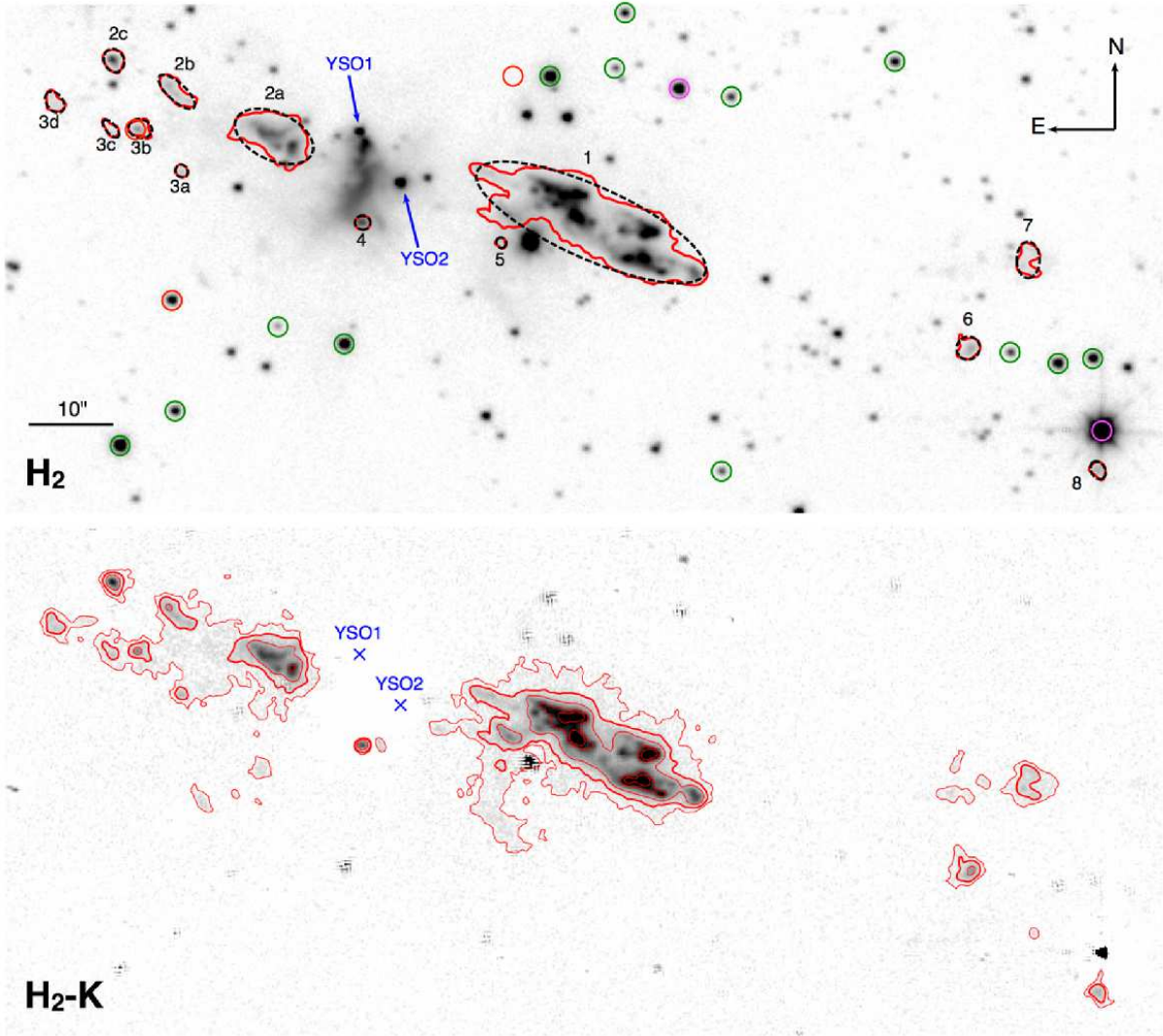}
\caption{Top: UKIRT/WFCAM \htwo\ image of the \core\ outflow. 
Red contours present the individual emission features defined by the threshold of 
three sigma above the background (thick contours in bottom panel), 
and black dashed ellipses are the results of the ellipse fitting 
of each contour.
Two YSOs at the center (YSO1 and YSO2) are marked.
Circles identify the mid-IR sources from GLIMPSE catalog/archive: red,
green, magenta colors indicate Class I, Class II including flat spectrum, 
and Class III YSOs defined by the spectral indices calculated between 2 and 8~\micron.
Bottom: The continuum-subtracted \htwo\ image.
Red contours are $1\sigma, 3\sigma, 10\sigma, 45\sigma$, and $80\sigma$
above the background. The thick red contours ($3\sigma$) present 
the threshold used for the identification of the \htwo\ outflow emission features.
\label{fig2}}
\end{center}
\end{figure*}

Figure~\ref{fig2} presents the UKIRT/WFCAM \htwo\ image of the \core\ outflow 
before (top) and after (bottom) continuum subtraction. The overall morphology 
of the outflow is bipolar but is composed of several discrete flows and 
knots. We identified the \htwo\ emission features of the outflow
to derive their geometrical parameters and \htwo\ line flux.
In the continuum-subtracted image, we estimated
the background value ($F_{bg}$) and determined a threshold for the outflow emission 
as three sigma above the background
($F_{bg} + 3\sigma \simeq 2.9 \times 10^{-20}~{\rm W\,m^{-2}}$).
In the bottom panel of Figure~\ref{fig2}, red contours are 1$\sigma$,
3$\sigma$, 10$\sigma$, 45$\sigma$, and 80$\sigma$ above the background, 
and the thick contours ($=F_{bg}+3\sigma$) present the threshold,
which is also drawn by red contours in the top panel of the figure.
In this process, 
we excluded artifacts and emission features with the area smaller 
than $<$0.25~arcsec$^{2}$ (i.e., the area of a circle with its diameter of 1$\arcsec$) 
considering that a typical full width half maximum of the stellar PSF of 
the UWISH2 data is $<$1$\arcsec$ \citep{ioannidis12a}.
In total, we identified 13 \htwo\ emission features and assigned the numbers 
from \no1 to \no8; for the flows/knots in the same direction, 
we grouped them and assigned the same numbers with different alphabets 
(e.g., from \no2a to \no2c, and from \no3a to \no3d) as shown in 
the top panel of Figure~\ref{fig2}.

Since the \htwo\ emission defined by the contours at the threshold has 
irregular shapes, we fitted the individual emission features by an ellipse to derive 
their geometrical parameters. For the fitting, 
we used the IDL procedure FIT\_ELLIPSE included in the Coyote IDL Program 
Libraries\footnote{\href{http://www.idlcoyote.com}{http://www.idlcoyote.com}}. 
The fitting results are drawn as black dashed ellipses in the top panel of 
Figure~\ref{fig2}, and the derived geometrical parameters are presented in 
Table~\ref{tbl1}.
The central coordinate, size, and orientation angle ($\psi_{\rm ellipse}$) 
of the \htwo\ emission features have been derived by adopting the center position, 
length of major axis, and orientation angle (from north to east) of the major axis of 
the fitted ellipses, respectively. The position angles (PAs) PA1 and PA2 have been measured 
by the angle (from north to east) of the central position of the emission features 
with respect to YSO1 and YSO2, respectively, because the driving source is not clearly
known (see Section~\ref{sec:pa}). 
Table~\ref{tbl1} also lists the area and \htwo\ line flux (see Section~\ref{sec:lum})
of the emission features that have been directly estimated from their contours.

\begin{deluxetable}{lcccccccccc}
\tabletypesize{\tiny}
\setlength{\tabcolsep}{0.07in}
\tablecaption{
Physical Parameters of the \htwo\ Emission Features of the \core\ Outflow \label{tbl1}}
\tablewidth{0pt}
\tablehead{
\colhead{ID} & \colhead{R.A. (J2000)} & \colhead{Dec (J2000)} & 
\colhead{Size} & \colhead{Size} & 
\colhead{$\psi_{\rm ellipse}$} & \colhead{PA1} & \colhead{PA2} &
\colhead{Area} & \colhead{Line Flux} & \colhead{UWISH2 Source ID} \\
\colhead{ } & \colhead{ } & \colhead{ } & 
\colhead{(arcsec)} & \colhead{(pc)} & 
\colhead{(deg)} & \colhead{(deg)} & \colhead{(deg)} &
\colhead{(arcsec$^{2}$)} & \colhead{($10^{-18}$ \wm)} & \colhead{ }  
} 
\startdata
1 & 19:29:15.68 & 17:56:12.6 &   59.0 &   0.97 &   65 & 
 69 &  78 &  179.2 & 1063.0 & UWISH2\_053.13615+0.07569 \\
2a & 19:29:18.31 & 17:56:22.7 &   19.7 &   0.32 &   65 & 
\nodata &  70 &   39.9 &  137.9 & UWISH2\_053.14447+0.06663 \\
2b & 19:29:19.11 & 17:56:28.0 &   11.1 &   0.18 &   47 & 
\nodata &  68 &    8.3 &   13.7 & UWISH2\_053.14447+0.06663 \\
2c & 19:29:19.63 & 17:56:31.7 &    5.7 &   0.09 &   25 & 
\nodata &  67 &    5.1 &   18.1 & UWISH2\_053.14796+0.06517 \\
3a & 19:29:19.07 & 17:56:18.6 &    3.1 &   0.05 &   42 & 
\nodata &  87 &    1.7 &    2.4 & UWISH2\_053.14447+0.06663 \\
3b & 19:29:19.42 & 17:56:23.6 &    5.6 &   0.09 &   73 & 
\nodata &  78 &    4.4 &    7.3 & UWISH2\_053.14447+0.06663 \\
3c & 19:29:19.66 & 17:56:23.6 &    5.1 &   0.08 &   40 & 
\nodata &  79 &    2.2 &    2.4 & UWISH2\_053.14447+0.06663 \\
3d & 19:29:20.12 & 17:56:26.8 &    5.6 &   0.09 &   37 & 
\nodata &  77 &    3.8 &    4.5 & UWISH2\_053.14447+0.06663 \\
4 & 19:29:17.57 & 17:56:12.6 &    3.7 &   0.06 &   80 & 
  2 & 137 &    2.4 &    7.9 & \nodata \\
5 & 19:29:16.43 & 17:56:10.2 &    2.5 &   0.04 &   35 & 
 52 &  59 &    1.2 &    1.1 & UWISH2\_053.13615+0.07569 \\
6 & 19:29:12.56 & 17:55:57.7 &    5.6 &   0.09 &  137 & 
 70 &  74 &    5.4 &    9.1 & UWISH2\_053.12637+0.08503 \\
7 & 19:29:12.06 & 17:56:08.1 &    8.6 &   0.14 &    0 & 
 79 &  83 &    7.6 &    8.3 & UWISH2\_053.12785+0.08817 \\
8 & 19:29:11.48 & 17:55:43.3 &    4.5 &   0.07 &   29 & 
 65 &  68 &    2.8 &    4.3 & UWISH2\_053.12064+0.08683 \\
\tableline
Total & \nodata & \nodata & 126\tablenotemark{\dag} &  1.04\tablenotemark{\dag} &
\nodata & 67\tablenotemark{*} & 74\tablenotemark{*} & 264 & 1280 & \nodata \\
\enddata
\tablecomments{R.A. (J2000), Dec (J2000) = central coordinates derived from
the center of the fitted ellipses; Size and $\psi_{\rm ellipse}$ = length and 
orientation angle (from north to east) of the major axis of the fitted ellipses; 
PA1 and PA2 = position angle (from north to east) of the emission features 
with respect to YSO1 and YSO2 (see Section~\ref{sec:pa}); 
Area = area of the individual contours; 
Line Flux = \htwo\ line flux directly measured from the individual contours with 
the uncertainty of $\sim$10\%; UWISH2 Source ID = from the UWISH2 extended
\htwo\ source catalog \citep{froebrich15}.}
\tablenotetext{\dag}{Estimated from the largest separation of the
individual emission features from \nosm2c to \nosm8 that are connected 
with a straight line by assuming YSO2 as a driving source.}
\tablenotetext{*}{Mean position angle, but \nosm4 is not included (see Section~\ref{sec:pa}).}

\end{deluxetable}


\subsubsection{Apparent Morphology}\label{sec:apparent}

The outflow can be divided into the main flow (from \no1 to \no5) and the faint 
knots (from \no6 to \no8) in the southwest. The main outflow has a bipolar shape along 
the NE-SW direction. While the NE flow is made up of two groups of flows (\no2 and \no3), 
the SW flow is identified as one flow (\no1) because 
the whole flow is brighter than the threshold.
This brightness difference between the two flows implies that the brighter SW flow 
is likely blueshifted if both flows originate from a single source. 

The flow \no1 is composed of several, at least six bright flows and knots
as shown by the contours at the higher levels than the threshold 
in the bottom panel of Figure~\ref{fig2};
the faint emission \no5 also can be a part of \no1. 
The sub-flows in \no1 have slightly different orientations and 
show a bow-shock-like feature at their tips (see Figure~\ref{fig8} for 
higher-resolution images).
The flow \no2a consists of two components: a compact knot and a flow 
with a bow-shock-like tip that is well connected to the flows \no2b and \no2c.
The emission features grouped as \no3 are smaller and fainter than 
those in \no2. As shown by the one-sigma level
contours in the bottom panel of Figure~\ref{fig2},
\no2 and \no3 have different orientations from the outflow center,
and \no3a is not well aligned with the other knots in \no3.
The complicated structure with several flows of different orientations 
seen in the flows \no1, \no2, and \no3 may imply 
multiple precessing jets; we discuss this possibility in detail 
in Section~\ref{sec:origin}.
The emission \no4 is near the center of the main flow, at the southern
end of the central nebula. 
Since \no4 is detected in both UKIRT and Subaru images,
it is not residual nebula emission from continuum subtraction 
but real \htwo\ emission.
The association between \no4 and the other \htwo\ features of the outflow 
is ambiguous because the direction from the central YSOs to \no4 is almost 
perpendicular to the whole outflow in the NE-SW direction, 
raising a question whether \no4 is a part of another, separate outflow 
(see Section~\ref{sec:origin}).

The remaining emission features \no6, \no7, and \no8 located in the southwest 
are faint but clearly seen in the continuum-subtracted image.
They are rather far away, but
there is no other YSO or other object that can emit \htwo\ emission,
indicating their association with the main flow.
We note that we have found faint emission features on the opposite
side as well, i.e., toward the northeast, outside the region shown in Figure~\ref{fig2}, 
at a similar distance to \no8 from the outflow center; however, it is unclear 
whether they are associated with \core\ because their surroundings are 
complicated, with another \htwo\ emission features and YSO candidates, so  
we do not include them in this study and only consider the \htwo\ emission
presented in Figure~\ref{fig2}, i.e., from \no1 to \no8.

\subsubsection{Size and Mass Ejection Frequency}\label{sec:size}

Size of the individual \htwo\ emission features of the outflow estimated 
from the major axis of the fitted ellipses is from 3$\arcsec$ to $\sim$60$\arcsec$ 
typically with the larger size for the brighter ones (Table~\ref{tbl1}).
The total length of the outflow is $\sim$80$\arcsec$ if only the main flow 
from \no1 to \no5 is considered, or $\sim$130$\arcsec$ if the faint knots 
in the southwest (\no6, \no7, \no8) are included, corresponding to
$\sim$0.7 and $\sim$1~pc, respectively, at the distance to IRDC G53.2, 1.7~kpc.
From the length of the one lobe (the SW lobe), from 0.35 to 0.74~pc, 
we constrain the dynamical age of the outflow, although it gives 
a wide range of timescales depending on the outflow velocity and inclination with 
respect to the plane of the sky: from 16,000 to 36,000~yr 
with the velocity of 20~\kms; from 3,000 to 7,200~yr with the velocity of 100~\kms. 
The assumed velocity range from 20 to 100~\kms\ is adopted from 
the observed proper motions of \htwo\ outflows \citep{khanzadyan03,raga13}, 
but we note that an outflow velocity can be as high as 150--300~\kms\ \citep[e.g.,][]{bally15}.
While protostellar outflows from low-mass YSOs are typically in sub-parsec scale 
with a small ($\sim$10\%) fraction of parsec scale 
outflows \citep{stanke02,davis08,davis09,ioannidis12a},
the outflows from high-mass YSOs tend to be more spatially extended \citep{varricatt10,caratti15}. 
Thus, the relatively large ($\sim$1~pc) size of the \core\ outflow suggests
that the outflow-driving source is likely massive.

As described above, the outflow is composed of several flows and knots.
The discrete components or clumpy features are often interpreted as episodic 
mass ejection \citep[][and references therein]{dunham14}, 
so we measured the separations between the emission
features that are well aligned in order to examine the mass ejection frequency. 
The separations between the knots in the main flow are typically around 
10$\arcsec$: the separations between \no2a and \no2b, between \no2b and
\no2c, and between \no3b and \no3d by assuming YSO2 as a driving source.
The separation between two sub-knots in \no1 ({\it sub2} and {\it sub5} in 
Figure~\ref{fig8}) along the line from YSO2 is also $\sim$10$\arcsec$.
The separation of 10$\arcsec$ corresponds to a time gap about 1,000 yr 
with the outflow velocity of 80~\kms.
(Here, we assume the outflow velocity to be the same as the velocity assumed in
two studies using the same UWISH2 data for comparison; \citealt{ioannidis12b}, \citealt{froebrich16}.)
This time gap of $\sim$1,000 yr is comparable to the typical time gaps 
between the \htwo\ knots of the outflows 
in Serpens/Aquila \citep[1,000--2,000~yr;][]{ioannidis12b} and 
Cassiopeia/Auriga \citep[1,000--3,000~yr;][]{froebrich16}.
The separations to the faint knots in the southwest are larger:
the separations between the southernmost sub-knot in \no1 ({\it sub1} in Figure~\ref{fig8})
and \no8 along the line from YSO1 is $\sim$60$\arcsec$;
the separation between the same knot and \no6 
along the line from YSO2 is $\sim$40$\arcsec$.
These large separations 
may imply that the distant, faint knots are not a part of the \core\ outflow or 
that we have missed much fainter emission between them, but it is also possible
that the mass ejection frequency and/or outflow velocity is not
constant over time or that multiple jets with different direction and velocity 
have been explosively ejected.

\subsubsection{Position Angle}\label{sec:pa}

We present the PAs of the \htwo\ emission features
in Table~\ref{tbl1}. Since the outflow-driving source is unknown, we separately 
measured PAs of the emission features with respect to YSO1 and YSO2 from 
north to east, and defined them as PA1 and PA2, respectively.
For YSO1, we only consider the emission in the southwest
(\no1, \no5, \no6, \no7, \no8) because the NE flow requires
a high degree of precession if it has been ejected from YSO1; 
for YSO2, we consider all of the emission features except \no4 
that has a different PA as described in Section~\ref{sec:apparent}.
The measured PAs of the emission features with respect to YSO1 (PA1) 
are from 52$\arcdeg$ to 79$\arcdeg$ with the mean of $67\arcdeg^{+12}_{-15}$,
and the PAs with respect to YSO2 (PA2) are from 59$\arcdeg$ to 87$\arcdeg$ with 
the mean of $74\arcdeg^{+13}_{-15}$.
The accurate PA of the entire outflow can be measured once the driving source
is confirmed, but our current results indicate that in any case, 
the PA will be around 70$\arcdeg$.
Table~\ref{tbl1} also shows that the PA of \no4 is indeed very different from 
those of the other features as expected---the estimated PAs are 2$\arcdeg$ and 
137$\arcdeg$ with respect to YSO1 and YSO2, respectively.

\subsection{\htwo\ Outflow Luminosity}\label{sec:lum}

We derived the \htwo\ luminosity of the \core\ outflow from the UWISH2 image.
We first estimated the \htwo\,1-0\,S(1) emission line flux ($F_{2.12}$)
given as $F_{2.12} = F_{0}({\rm DN}/t_{\rm exp})10^{-0.4ZP}$ from
the continuum-subtracted image,
where $F_{0}$ ($ = 9.84 \times 10^{-12}$~\wm)
is a total in-band flux of the \htwo\ filter,
DN is the total sum of pixel values of the region of interest,
$t_{\rm exp}$ is the exposure time ($=60$~s), 
and $ZP$ is the zero-point magnitude ($=21.125$~mag for our image) 
written in the image header.
Calculating the total sum of pixel values, we multiplied a factor of 1.10 
in order to compensate the \htwo\ line flux that is included in the {\it K}-band image 
so subtracted during the continuum-subtraction process (Lee, Y.-H. et al. in preparation).
With the uncertainty of $\sim$10\% in flux measurements,
the estimated 2.12~\micron\ line flux of each contour determined in 
Section~\ref{sec:ident} is from $1.1 \times 10^{-18}$ 
to $1063.0 \times 10^{-18}$ \wm\ as listed in Table~\ref{tbl1}, giving the total 
line flux of $1.28 \times 10^{-15}$ \wm\ for the total area of 264 arcsec$^{2}$.
We note that the threshold we used to identify the \htwo\ emission 
features, three sigma above the background,
is rather conservative, so our flux estimation gives a lower limit.
For comparison,
the \core\ outflow is also included in the UWISH2 extended \htwo\ source 
catalog \citep{froebrich15} as we present the UWISH2 source IDs 
in the last column of Table~\ref{tbl1}.  
The contours of the UWISH2 sources corresponding 
to the \core\ outflow are almost consistent with the contours at the level 
of one sigma above the background in Figure~\ref{fig2} (bottom), enclosing 
the area about three times larger than our results.
The different threshold values, however, insignificantly affect the total 
line flux because most of the additional area have very low surface brightness.
The \htwo\ line flux of the \core\ outflow region from the UWISH2 
catalog \citep{froebrich15} is $\sim$10\% larger than our estimation.

From the total 2.12~\micron\ line flux 
$F_{\rm 2.12,obs} \sim 1.28 \times 10^{-15}$ \wm,
the 2.12~\micron\ luminosity of the outflow at 
the distance of 1.7~kpc is $L_{\rm 2.12,obs}\sim 0.1~L_{\sun}$, 
but this is highly underestimated because
extinction toward IRDC cores is expected to be large.
Since the extinction of \core\ has not been previously measured, 
we constrain the lower and upper limits using the optical depth 
of the {\it Spitzer} dark cloud SDC053.158+0.068 \citep{peretto09} and 
the $^{13}$CO column density $N(^{13}{\rm CO})$ obtained from 
the $^{13}$CO $J=$~1--0 data in the Boston University-Five College Radio 
Astronomy Observatory Galactic Ring Survey \citep[GRS;][]{jackson06}, respectively.
SDC053.158+0.068 is a large (major axis $\sim 300\arcsec$) dark cloud 
that includes \core. The averaged optical depth of SDC053.158+0.068 measured
at 8~\micron\ is 0.68, or $A_{\rm 8\mu m} = 0.63$~mag \citep{peretto09}.
The extinction $A_{\rm 8\mu m} = 0.63$~mag is converted to 
$A_{K} \sim 1.5$~mag or $A_{\rm V} \sim 15$~mag by the mid-IR extinction curves 
derived in \citet{flaherty07} and \citet{chapman09}; this value $A_{\rm V} \sim 15$~mag
can be the lower limit of the extinction of \core, a denser core inside the dark cloud.
In our previous study, we derived the $N(^{13}{\rm CO})$ map of IRDC G53.2
from the GRS $^{13}$CO $J=$~1--0 data \citep{kim15}. 
In the GRS column density map with a large angular resolution (46$\arcsec$) 
and pixel scale (20$\arcsec$), \core\ is covered by a few pixels 
with $N(^{13}{\rm CO})$ around $9 \times 10^{16}$~cm$^{-2}$. 
From $N(^{13}{\rm CO})$, we derive $N_{\rm H_2}$ assuming the same numbers of 
$^{12}{\rm CO}/^{13}{\rm CO}=60$ \citep[Equation (3) of][]{milam05}
and $n(^{12}{\rm CO})/n({\rm H_2}) = 1.1 \times 10^{-4}$ \citep{pineda10} 
used to derive the $N(^{13}{\rm CO})$ map \citep[Section~2 of][]{kim15}.
The derived $N_{\rm H_2}$ is $\sim 5 \times 10^{22}$~cm$^{-2}$, 
or $A_{\rm V} \sim 50$~mag.
Since the extinction value derived from $N(^{13}{\rm CO})$ takes the entire thickness
of the molecular cloud along the line of sight into account, we adopt $A_{\rm V}= 50$~mag 
as the upper limit of the extinction of \core.
From the above, the extinction of \core\ is 15~mag $< A_{\rm V} <$~50~mag, leading
to the extinction-corrected 2.12~\micron\ luminosity of the outflow $0.4 < L_{\rm 2.12}/L_{\sun} < 10$.
Then, we derive the total \htwo\ luminosity (\lhtot) applying the ratio between 
the 2.12~\micron\ intensity ($I_{2.12}$) and the total \htwo\ intensity ($I_{\rm H_2}$).
While the assumption $I_{2.12}/I_{\rm H_2} \sim 0.1$ is commonly 
used \citep[e.g.,][]{stanke02,caratti06,ioannidis12b}, $I_{2.12}/I_{\rm H_2}$ is in fact 
a function of gas temperature in LTE conditions. We assume the gas temperature 
of 1,500--3,000~K, although the temperatures of outflows from high-mass YSOs tend to be 
relatively higher \citep[$\sim$2,500~K;][]{smith97,davis04,caratti15}, 
and apply $I_{2.12}/I_{\rm H_2}$ of 0.1--0.05 \citep[Figure~3 of][]{caratti06}. 
The \lhtot\ of the \core\ outflow finally derived in the constrained ranges of extinction 
and temperature is, therefore, $6 \pm 2 < L_{\rm H_2}/L_{\sun} < 150 \pm 50$.

In several studies,
the \htwo\ luminosity of outflows shows a strong correlation with the bolometric 
luminosity (\lbol) of the driving sources \citep[e.g.,][]{caratti06,cooper13,caratti15}, 
thus we can constrain the driving source of the \core\ outflow 
from the derived outflow luminosity.
The luminosities of the outflows driven by low-mass YSOs are typically
lower than the luminosity of the \core\ outflow.
For example,
\lhtot\ of 23 protostellar jets driven by low- and intermediate-mass
YSOs studied in \citet{caratti06} ranges from 
0.007 to 0.76~$L_{\sun}$; 
the outflows detected in Serpens/Aquila from 
the UWISH2 survey show $L_{\rm H_{2},obs}$ ranging
from 0.01 to 1.0~$L_{\sun}$,
which is less than a few solar luminosity after extinction correction by using
a typical extinction of the region, $A_{K}=1$~mag \citep{ioannidis12b}.
The driving source of the \core\ outflow, therefore, is expected to
be a high- or at least intermediate-mass YSO.
We further constrain \lbol\ of the driving source by adopting 
the empirical relationship between \lhtot\ of the outflows and \lbol\ of
the protostars derived from the excitation conditions and visual extinction 
values obtained by spectroscopic observations, 
a relationship that is defined as 
$L_{\rm H_2} \propto L_{\rm bol}^{\alpha}$ with
$\alpha=0.59$ or $\alpha=0.57 \sim 0.62$ for outflows from very young 
(Class 0 and Class I) low-mass or high-mass YSOs, 
respectively \citep[Figure~9 of][]{caratti15}.
On the relation of $L_{\rm H_2} \propto L_{\rm bol}^{\alpha}$ 
with $\alpha \sim 0.6$, the \lbol\ of the driving source expected from 
the outflow luminosity $6 \pm 2 < L_{\rm H_2}/L_{\sun} < 150 \pm 50$ is
$\sim 10^4 < L_{\rm bol}/L_{\sun} < 10^6$, supporting a high-mass protostar 
as a driving source of the \core\ outflow.

Although the rough information on the environmental conditions 
such as visual extinction or gas temperature provides 
a wide range of the \htwo\ luminosity of the \core\ outflow, 
the constrained \lbol\ suggests that 
the \core\ outflow is likely driven by a high-mass YSO.
However, we note that we have assumed a single source origin
for the entire \htwo\ emission in the above discussion, leaving  
a possibility of multiple outflow-driving YSOs with lower luminosity/mass, 
which will be discussed in Section~\ref{sec:origin}.

\section{Search for \feii\ Emission in \core}\label{sec:feii}

\feii, together with \htwo, is one of the prominent emission lines tracing protostellar jets.
In outflows/jets with \htwo\ emission, \feii\ emission, in particular 
the \feii\ 1.644~\micron\ lines in near-IR, are frequently observed as well
regardless of mass and evolutionary stage of the exciting stars
\citep[e.g.,][]{reipurth00,nisini02,giannini04,caratti06,caratti15,cooper13},
although the detection rates, morphologies, and spatial distributions
are different because these two lines arise from different shock origins:
\htwo\ lines trace slow and non-dissociative shocks whereas
\feii\ lines trace fast and dissociative shocks \citep{nisini02,hayashi09}.
Since the \core\ outflow is strong and well-defined by a bipolar shape 
in \htwo, it can be expected that \feii\ emission is also observed as 
a narrow jet emitted from the central YSOs as seen in a number of
Herbig-Haro (HH) objects \citep[e.g., HH~300, HH~111;][]{reipurth00}
or as compact knots (e.g., HH~223: \citealt{lopez10}; G35.2N: \citealt{lee14}).

\begin{figure}
\begin{center}
\includegraphics[width=0.87\textwidth]{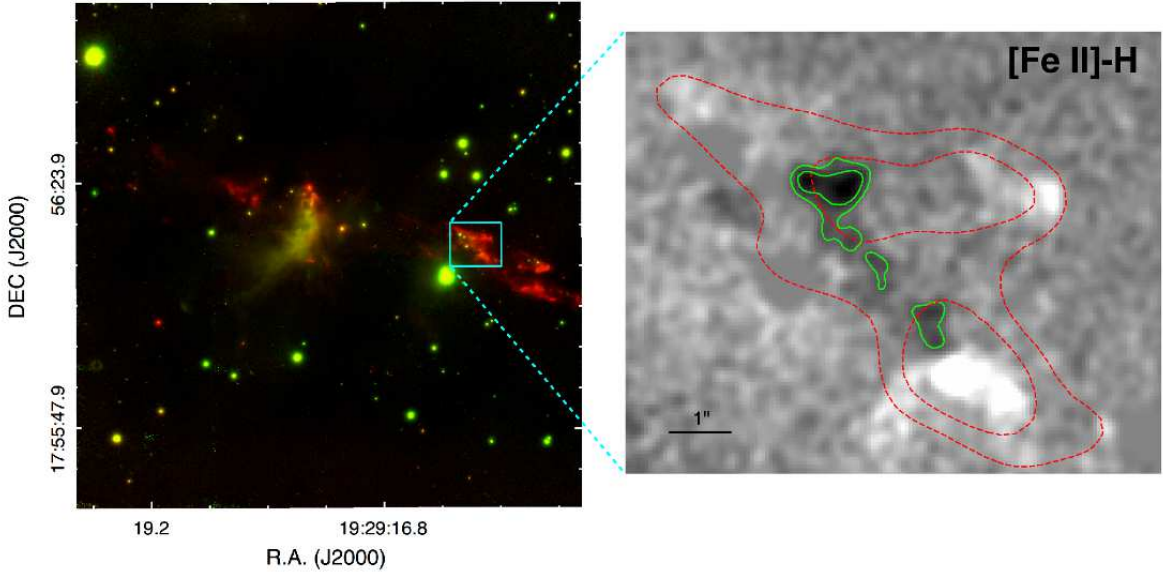}
\caption{Left: Subaru/IRCS image of \core. Red is \htwo, and green is \feii.
The cyan box marks the region where the \feii\ emission features are detected.
Right: The continuum-subtracted \feii\ image of the cyan-box region 
in the inverted-gray color scale. The image is smoothed by using a Gaussian
function, and the residual features from point source subtraction are masked to clearly 
show the \feii\ emission. Green contours are the detected \feii\ emission with 
the contour levels of $3\sigma$ ($5 \times 10^{19}~{\rm W\,m^{-2}\,arcsec^{-2}}$)
and $6\sigma$ ($9 \times 10^{19}~{\rm W\,m^{-2}\,arcsec^{-2}}$) above the background
estimated from the smoothed image.
Red dashed contours are the \htwo\ emission ($45\sigma$ and $80\sigma$ above
the background in Figure~\ref{fig2}) drawn for comparison. 
The negative extended features shown in white 
are residual \htwo\ line emission included in {\it H}-band that have 
remained after continuum subtraction.
\label{fig3}}
\end{center}
\end{figure}

We searched for \feii\ 1.644~\micron\ emission associated with the \core\ outflow. 
We found no \feii\ emission in the UWIFE image with a typical root-mean-square 
noise level of $8.1 \times 10^{-20}$~${\rm W\,m^{-2}\,arcsec^{-2}}$ \citep{lee14}
but detected faint emission features in the Subaru/IRCS image owing to 
the higher sensitivity. As the Subaru/IRCS images in Figure~\ref{fig3} show,
the \feii\ emission was found around the sub-flows 
in the \htwo\ flow \no1 ({\it sub4} and {\it sub5} in Figure~\ref{fig8}). 
The right panel of Figure~\ref{fig3} is the continuum-subtracted \feii\ image 
in the inverted-gray color scale with the \feii\ emission drawn by green contours.
Since the background is very noisy and the \feii\ emission features are barely
seen even in the continuum-subtracted image,
we smoothed the image by a Gaussian function with three pixels.
In the figure, the green contours represent three- and six-sigma above the background 
estimated from the smoothed image, the negative features seen in white are 
the \htwo\ lines included in {\it H}-band that have remained after continuum 
subtraction, and red dashed lines are the \htwo\ 2.12~\micron\ contours
drawn for comparison. 

The detected \feii\ emission is very small with the total length of 
$\sim$3$\arcsec$ (or the area of $\sim$1~arcsec$^{2}$) and faint;
the three-sigma flux of the \feii\ line is $\sim 5.7 \times 10^{-19}$ \wm\
with the uncertainty of 10\%, or the surface brightness is 
$\sim 5.6 \times 10^{-19}$ ${\rm W\,m^{-2}\,arcsec^{-2}}$, which is
$<$10\% of the surface brightness of the \htwo\ emission 
estimated as $\sim 4.8\times10^{-18}$ ${\rm W\,m^{-2}\,arcsec^{-2}}$ 
from the total \htwo\ line flux and area (Table~\ref{tbl1}).
Although the \feii\ emission is spatially coincident with the \htwo\ emission, 
it is difficult to conclude that the \feii\ emission is associated with the \htwo\ outflow 
because \feii\ knots are generally expected to be observed at the tips of 
the \htwo\ bow shocks, where the shock velocities are high and 
the gas will be dissociated, rather than behind 
the \htwo\ bow shocks \citep[e.g.,][]{davis99,davis00,lopez10,lee14,bally15}.

Detection of further \feii\ emission in \core\ with high signal-to-noise ratio requires 
deeper imaging observations, but the marginal detection of \feii\ emission can be 
interpreted as intrinsically fainter or absent \feii\ emission compared to \htwo\ emission.
In the outflows from high-mass YSOs, 
the \feii\ detection rate with respect to the \htwo\ detection rate tends 
to be low, $\lesssim$50\% or much less \citep{cooper13,wolf13,caratti15}; 
the brightness of \feii\ lines also tend to be weaker than the brightness
of \htwo\ lines \citep{caratti15}.
This can be attributed by different extinction effect between the two lines, 
but more likely, it is because
\htwo\ and \feii\ emission arise from different physical conditions 
such as gas density, temperature, or shock velocity. 
In the \core\ outflow, the strong, extended \htwo\ emission with 
very weak or negligible \feii\ emission may imply that slow, C-type shocks 
are dominant in \core. Further spectroscopic observations will 
be necessary in order to derive the physical conditions of environment
and confirm shock properties.

\section{Central Young Stellar Objects}\label{sec:yso}

The core \core\ is bright from IR to millimeter with large IR-excess emission. 
The central star (YSO1) was previously identified 
in the Red MSX Source (RMS) survey with the bolometric luminosity 
of (3--4)$\times10^3~L_{\sun}$ \citep[G053.1417+00.0705;][]{mottram11,lumsden13},
but it is in fact composed of two sources, YSO1 and YSO2, separated 
by $\sim$8$\arcsec$ in the {\it Spitzer} mid-IR images with higher spatial resolution.
Both YSOs are saturated in the MIPS 24~\micron\ image, but their SEDs 
with strong excess in mid-IR and spectral indices derived by using 
the available photometry from the GLIMPSE and MSX catalogs classify 
them as Class I YSOs that have a dusty envelope
infalling onto a central protostar \citep{kim15}. 
The two YSOs are observed in near-IR wavebands as well.
Both are fairly bright in {\it K}-band, marginally detected in {\it H}-band  
but not observed in {\it J}-band, indicating that they are deeply embedded.
The evolutionary stages of the YSOs, with the proximity to 
the center of the outflow (see Figure~\ref{fig2}), suggest that one of them or both 
can be the driving source of the \core\ outflow. 
The coordinates of YSO1 and YSO2 are
($\alpha_{2000},\delta_{2000}$)=
(${\rm 19^{h}29^{m}17\,\fs60, +17\arcdeg56\arcmin23\farcs3}$) 
and ($\alpha_{2000},\delta_{2000}$)=
(${\rm 19^{h}29^{m}17\,\fs26, +17\arcdeg56\arcmin17\farcs3}$), respectively. 
Below, we will discuss  their photometric variability in near-IR
and physical parameters constrained from SED analysis.

\subsection{Near-IR Photometric Variability}\label{sec:phot}

\begin{figure}
\includegraphics[width=1\textwidth]{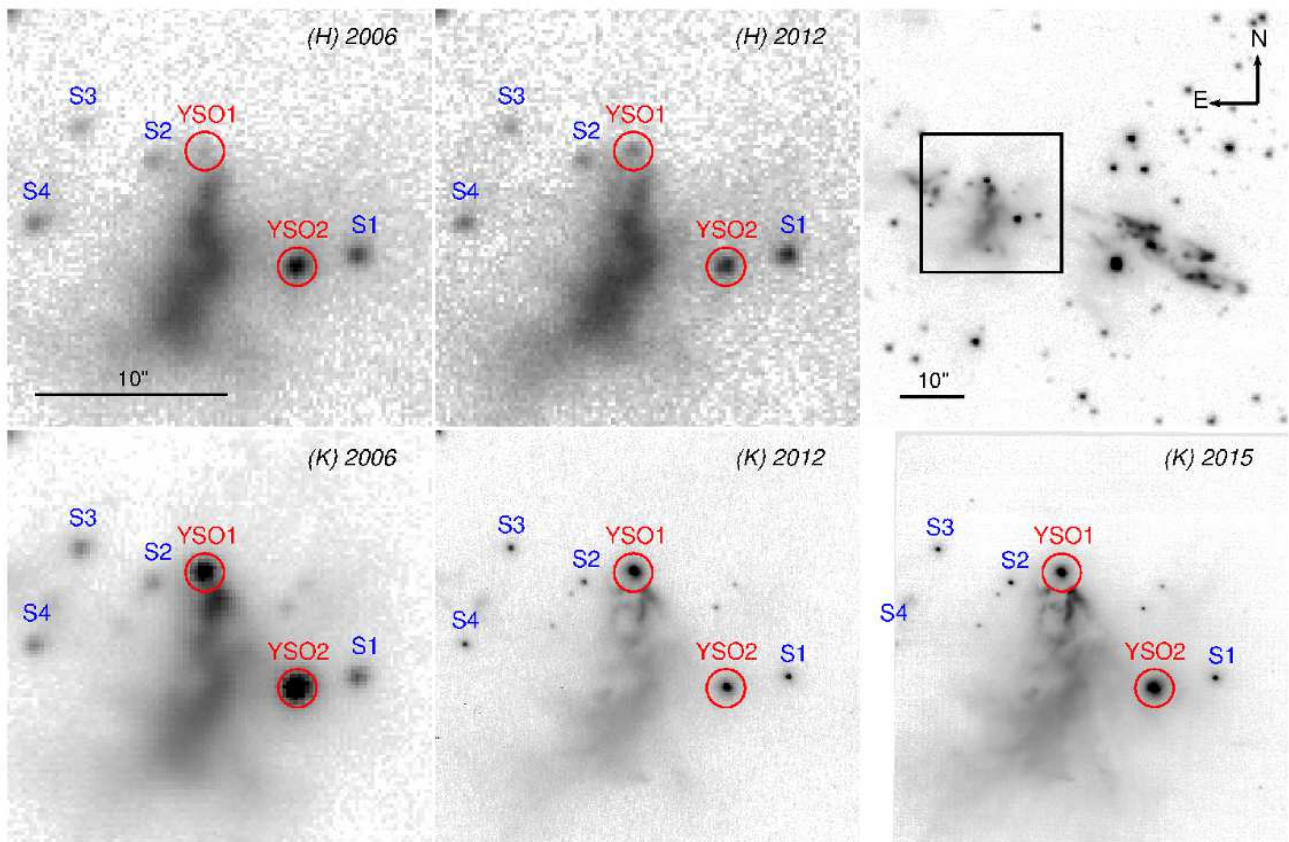}
\caption{{\it H}- and {\it K}-band images of the central part of 
\core\ obtained at different epochs. The \htwo\ image from UWISH2
at the rightmost in upper panel shows the region of interest with 
a black box. In the rest five figures, the observed bands and 
years are presented in the right upper side. 
Two {\it H}-band images and the 2006 {\it K}-band image
were obtained by UKIRT, whereas the {\it K}-band images 
in 2012 and 2015 were obtained by Subaru/IRCS and Gemini/NIRI, respectively. 
YSO1, YSO2, and the reference stars (from S1 to S4) used for 
differential photometry are also marked.
\label{fig4}}
\end{figure}

YSOs are known to commonly show variability 
\citep[e.g.,][]{carpenter01,carpenter02,morales11,johnstone13,wolk13,rebull15}.
Since we have several {\it H}- and {\it K}-band images of the central part 
of \core\ obtained at different epochs between 2006 and 2015, 
we compare the brightness of YSO1 and YSO2 over time.
We exclude the 2MASS images in which both YSOs are not clearly
resolved and likely contaminated by bright emission of the extended,
central nebula due to low resolution.
In {\it H}-band, 
we have the UKIRT images taken in 2006 and 2012, and 
the Subaru/IRCS image taken in 2012; in {\it K}-band, 
we have the UKIRT, Subaru/IRCS, and Gemini/NIRI images 
obtained in 2006, 2012, and 2015, respectively. 
Both the UKIRT and Subaru {\it H}-band images in 2012
were obtained in July, so we only use the UKIRT image 
for consistency with the 2006 data.
The images are compared in Figure~\ref{fig4}. 
The Subaru and Gemini images with higher resolution show 
a more complex structure of the central nebula, and variations in 
relative brightness between YSO1 and YSO2 are seen in 
some images, e.g., the {\it K}-band images between 2012 and 2015. 

\begin{figure}
\begin{center}
\includegraphics[width=0.5\textwidth]{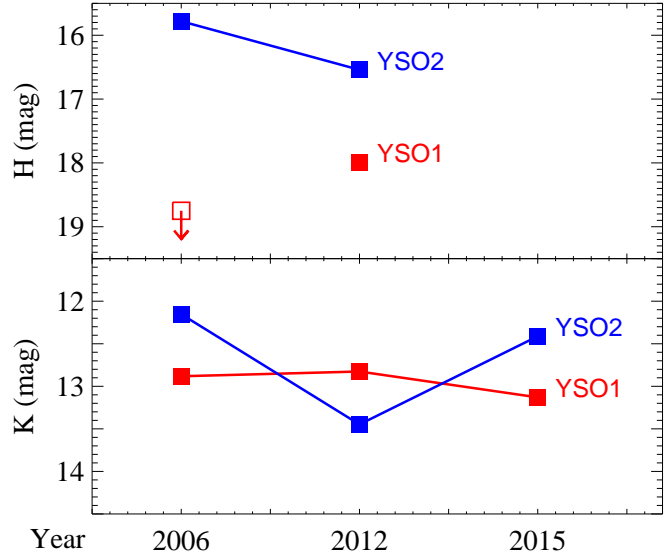}
\caption{{\it H}- and {\it K}-band magnitudes of YSO1 and 
YSO2 over time. Open symbol with an arrow presents 
the upper limit of YSO1 (=18.75~mag) in the UKIRT {\it H}-band image adopted
from the typical 90 per cent completeness limit of UKIDSS GPS 
estimated in uncrowded fields \citep{lucas08}.
Photometric errors ($\lesssim 10\%$) are smaller
than the symbol size.
\label{fig5}}
\end{center}
\end{figure}

We estimated the flux of YSO1 and YSO2 from each image.
For the UKIRT images, we performed PSF photometry of the point sources 
using \textsc{starfinder} \citep{diolaiti00} based on the 2MASS catalog \citep{2mass}. 
For the Subaru and Gemini images, we applied differential photometry 
using the point sources identified in the UKIRT images because our interest is 
photometric variability of the YSOs. We used four stars without IR excess 
marked in Figure~\ref{fig4} as reference stars.
Table~\ref{tbl2} lists the estimated magnitudes of YSO1, YSO2, and
the reference stars, and Figure~\ref{fig5} compares
the magnitudes of the two YSOs over time. 
Photometric errors from \textsc{starfinder} are negligibly small but 
highly underestimated because it only accounts for the errors 
from PSF fitting and does not include other possible uncertainties 
such as the uncertainty from background variations 
that mostly contribute to photometric uncertainties, 
in particular around the region with nebula emission. 
In Table~\ref{tbl2}, the magnitudes of the reference stars 
at different epochs show the uncertainties less than or around 10\%,
so we adopt the photometric errors of $\lesssim$10\%.
We note that S2 exceptionally shows a large difference 
of $\sim$25\% between 2006 and 2012/2015 in {\it K}-band.
This large uncertainty is likely because S2 is located
so close to the nebula that it is more affected by
the extended nebula emission particularly in the UKIRT image 
with lower resolution than in the other two images;
if the PSF baseline of S2 in the UKIRT image 
were determined on the level of the nebula emission,
the source flux could have been underestimated from the 
higher baseline, resulting in the fainter brightness of S2.

Table~\ref{tbl3} presents the amplitudes of variability in
YSO1 and YSO2 between two time durations: from 2006 
to 2012 and from 2012 to 2015.
While the variability of YSO2 is obvious in both time durations 
with magnitude differences of $\gtrsim$1 mag in 
both {\it H}- and {\it K}-bands, 
the variability of YSO1 is rather ambiguous.
In {\it H}-band,
YSO1 was not detected in 2006 but appeared in 2012, giving
the magnitude difference larger than 0.76~mag from the detection 
limit 18.75~mag of the UKIRT {\it H}-band image \citep{lucas08};
in {\it K}-band, however, 
YSO1 maintained its brightness within the photometric
uncertainty between 2006 and 2012.
This discrepancy can be also explained by 
the contamination from the central nebula,
but in a manner opposite to S2, the nebula emission could have been
included in the source flux since YSO1 is located at the 
tip of the nebula as seen in Figure~\ref{fig4}, leading to
overestimation of the {\it K}-band flux of YSO1 in 2006.
It is less probable that the {\it H}-band flux is over/underestimated 
because the nebula emission in {\it H}-band is not as strong
as in {\it K}-band. We also note that the flux of YSO1 estimated
from the Subaru {\it H}-band image agrees well with that of the 
UKIRT image in the same year 2012.
Since the two {\it H}-band images in Figure~\ref{fig4} were obtained 
with the same telescope and the same instrument, 
we believe that the {\it H}-band magnitudes
in Table~\ref{tbl2} are reliable.
If assuming that the {\it K}-band flux in 2006 is 
overestimated, YSO1 in 2006 could have been fainter than 
presented in Table~\ref{tbl2} and become brighter in 2012,
consistent with
the photometric behavior in {\it H}-band.
The variability of YSO1 is also supported by the brightness
change between the 2012 and 2015 {\it K}-band images in which
the nebula contamination is likely insignificant owing to their 
higher resolution.

\begin{deluxetable}{ccccccccc}
\tabletypesize{\scriptsize}
\tablecaption{
{\it H}- and {\it K}-band Magnitudes of 
YSO1, YSO2, and Reference Stars \label{tbl2}}
\tablewidth{0pt}
\tablehead{
\colhead{}   &\colhead{} & \colhead{} &  
\multicolumn{2}{c}{{\it H}-band (mag)} &   \colhead{}   &
\multicolumn{3}{c}{{\it K}-band (mag)} \\
\cline{4-5} \cline{7-9} \\
\colhead{} &\colhead{R.A. (J2000)} & \colhead{Dec(J2000)} & 
\colhead{UKIRT2006} & \colhead{UKIRT2012}  & \colhead{}  & 
\colhead{UKIRT2006} & \colhead{Subaru2012} & \colhead{Gemini2015}}
\startdata
YSO1\tablenotemark{\dag} & 
19:29:17.60 & 17:56:23.3 & $>$18.75\tablenotemark{a} & 17.99 & & 12.88 & 12.83 & 13.13 \\
YSO2\tablenotemark{\dag} & 
19:29:17.26 & 17:56:17.3 & 15.78    & 16.54 & & 12.15 & 13.45 & 12.41 \\
\\
 S1 & 19:29:17.04 & 17:56:17.8 &  16.51 & 16.54 & & 14.35 & 14.34 & 14.34 \\
 S2 & 19:29:17.79 & 17:56:22.8 & 17.91 & 17.92 & & 15.47 & 15.23 & 15.24 \\
 S3 & 19:29:18.05 & 17:56:24.5 &  18.06 & 18.20 & & 14.94 & 15.09 & 15.13 \\
 S4 & 19:29:18.22 & 17:56:19.5 &  17.26 & 17.44 & & 15.04 & 15.10 & \nodata\tablenotemark{b} \\
\enddata

\tablenotetext{\dag}{YSO1 and YSO2 are the same as No. 1 and 2 in Table~3 of \citet{kim15}.
We note their coordinates are slightly different  because the coordinates in \citet{kim15} 
are adopted from the 2MASS catalog while the coordinates presented in this table are 
obtained from the UKIRT data.}
\tablenotetext{a}{YSO1 is not detected in the UKIRT {\it H}-band image in 2006. 
{\it H}=18.75 mag is the typical 90 per cent completeness limit of UKIDSS GPS 
estimated in uncrowded fields \citep{lucas08}.}
\tablenotetext{b}{S4 is out of the FOV of the Gemini image.}
\tablecomments{Photometric errors are $\lesssim$10\%.}

\end{deluxetable}


\begin{deluxetable}{ccccc}
\tabletypesize{\scriptsize}
\tablecaption{Amplitudes of Variability and Colors of YSO1 and YSO2 \label{tbl3}}
\tablewidth{0pt}
\tablehead{
\colhead{} & \colhead{$\Delta H_{\rm 2012-2006}$} & 
\colhead{$\Delta K_{\rm 2012-2006}$} &
\colhead{$\Delta K_{\rm 2015-2012}$} & 
\colhead{$\Delta [H-K]_{\rm 2012-2006}$}}
\startdata
YSO1 & $<-0.76$ & $-0.05$ & 0.3 & $<-0.71$ \\
YSO2 & 0.76 & 1.3 & $-1.04$ & $-0.54$ \\
\enddata
\end{deluxetable}


Flux measurement confirms the variability of both YSOs 
with the variances up to 0.76~mag in {\it H}-band for YSO1 and 
1.3~mag in {\it K}-band for YSO2.
Although the limited data only obtained at two or three epochs 
are not enough to find either the variability period or the full variability
amplitude, the observed variances of $\sim$1~mag give some 
implications on the variable characteristics.
There are several mechanisms that can produce variability
in YSOs: cold or hot spots on the stellar photosphere; 
changes in disk structure such as the location of the inner disk 
boundary, variable disk inclination, and changes in the accretion rate;
and variable extinction along the line of sight
\citep[][and references therein]{alves08,wolk13,contreras17b}. 
While most of these mechanisms are expected to make
relatively small variability amplitudes 
of $\Delta K < 1$~mag \citep[Table 6 of ][]{wolk13},
\citet{contreras17b} argued that 
the mechanisms such as variable extinction or changes in accretion 
rate can contribute to larger variability amplitudes if YSOs are
deeply embedded or experience a sudden increase of accretion
rate as the FU Orionis objects (FUors). 
Previous observations of YSOs in $\rho$ Oph and 
the Cyg OB7 region indeed show typical variability amplitudes
in {\it K}-band ranging from 0.01 to 0.8~mag and from 0.25 to 1.0~mag, 
respectively \citep{alves08, wolk13}, but larger variability 
amplitudes ($\Delta K > $1--2~mag) have been also found 
from a small number of YSOs in Cyg OB7 \citep{wolk13} and
from more than 400 YSOs identified in 119~deg$^{2}$ of the Galactic 
midplane by the VISTA Variables in the Via Lactea (VVV) survey,
which have been classified as eruptive variable YSOs \citep{contreras17b}.

The brightness changes of $\sim$1~mag observed  in YSO1 and 
YSO2 (Table~\ref{tbl3}) imply that they can be the candidates  
of eruptive variable YSOs with high variability amplitudes. 
Although the amplitude of YSO1, $\Delta H_{\rm 2012-2006}$,
is not large enough to satisfy the criterion of 
high amplitude ($\Delta K > 1$~mag) defined in \citet{contreras17b}, 
it only represent the amplitude between two epochs, giving a lower
limit of the full variability.
We compare $\Delta H_{\rm 2012-2006}$ and $\Delta [H-K]_{\rm 2012-2006}$
of YSO1 and YSO2 with colors and magnitudes of the variable YSOs
in the VVV survey. On the $\Delta (H-Ks)$ vs. $\Delta H$ 
plot \citep[Figure~21 of][]{contreras17b},
YSO1 falls in the ``bluer when brightening" quadrant with 
$\Delta H_{\rm 2012-2006} < -0.76$ and $\Delta [H-K]_{\rm 2012-2006} < -0.71$, 
and YSO2 falls in the ``bluer when fading" quadrant with
$\Delta H_{\rm 2012-2006}  = 0.76$ and $\Delta [H-K]_{\rm 2012-2006} = -0.54$.
Most of the VVV sources are elliptically distributed 
in a broad range passing the ``bluer when brightening" 
and ``redder when fading" quadrants
regardless of their types defined by the light curve morphology
\citep{contreras17b}. 
YSO1 follows this overall distribution. It cannot be determined 
whether YSO1 is an eruptive YSO, but YSO1 is clearly distinguished 
from the eclipsing binaries that are clustered around the origin.
YSO2 is a little apart from the overall elliptical distribution and located
in the ``bluer when fading" quadrant. In this region, the ones classified 
as faders that show a continuous decline in magnitude 
during the observed period \citep{contreras17b} are dominant
but with some eruptive YSOs as well. 
YSO2 cannot be a fader because it has become brighter again 
in 2015 but is possible to be an eruptive YSO.

The observed near-IR variability of $\gtrsim$1~mag together with 
the discrete features in the \htwo\ outflow (Section~\ref{sec:size})
suggest that YSO1 and/or YSO2 are the candidates of eruptive 
variable YSOs and may be the massive counterparts of MNors,
a newly proposed class of eruptive YSOs with the outburst 
duration between FUors and EXors \citep{contreras17a}.
Further consecutive observations to derive the full light curves 
and variability characteristics will be necessary to confirm 
this possibility.

\subsection{Spectral Energy Distribution Analysis}\label{sec:sed}

Both YSO1 and YSO2 have been classified as Class I by the spectral indices 
\citep[$\alpha = d~{\rm log}\,(\lambda\,F_{\lambda})/d~{\rm log}\,(\lambda)$;][]{lada87}
derived from their SEDs between 2 and 22~\micron\ (YSO1) or 
between 2 and 8~\micron\ (YSO2): $\alpha_{\rm YSO1}=1.88\pm0.62$ and 
$\alpha_{\rm YSO2}=2.21\pm0.13$ \citep{kim15}.
While spectral index, only determined by the SED shapes, can provide a way
to estimate the evolutionary stages of YSOs in a statistical sense if the sample
number is large enough
as discussed in \citet{kim15} as well as in other previous 
studies \citep[e.g.,][]{robitaille06,robitaille07},
it may not be appropriate to examine an individual source because 
the SED shapes can be affected by the inclination of the source to the line of sight 
or extinction toward the source \citep{robitaille07,forbrich10}; thus,
we fitted the SEDs of the two YSOs 
using the Python SED Fitter\footnote{\href{http://sedfitter.readthedocs.io/en/stable/index.html}
{http://sedfitter.readthedocs.io/en/stable/index.html}} (version 1.0)
to confirm their evolutionary stages and constrain the physical parameters 
based on physical models.
The SED Fitter developed by \citet{robitaille07} was previously available 
either in a command-line version or 
in an online version\footnote{\href{http://caravan.astro.wisc.edu/protostars/}
{http://caravan.astro.wisc.edu/protostars/}} but recently 
has been built in Python by the developer.
This fitting tool uses a large set of pre-calculated model SED 
grid \citep{robitaille06} made with the radiation transfer 
code from \citet{whitney03a, whitney03b}; the models were computed 
with 20,000 sets of parameters and 10 different viewing angles 
for each model set, i.e., 200,000 models in total. The model
SEDs are convolved with common filter bandpasses that are
available in the code or manually given by a user,
and the convolved fluxes
are fitted with the observed fluxes given as input data. 
In the fitting,
distance to the source and foreground extinction are allowed to be 
free parameters, and each fit is characterized by a chi-square 
value \citep{robitaille07}.

\begin{figure}
\begin{center}
\includegraphics[width=0.5\textwidth]{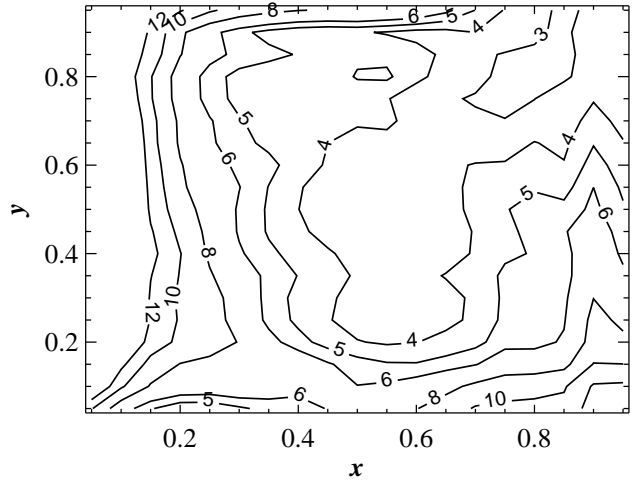}
\caption{Reduced chi-square (total chi-square divided by the number 
of data points) contours from the SED fitting of YSO1 and YSO2
to determine the relative fraction factors of the WISE 22~\micron\ flux ($x$) and 
the PACS 70~\micron\ flux ($y$). $x$ and $y$ are the fraction factors of YSO1, i.e.,
the fraction factors of YSO2 are $(1-x)$ and $(1-y)$.
\label{fig6}}
\end{center}
\end{figure}

Table~\ref{tbl4} lists mid- and far-IR fluxes of YSO1 and YSO2 used 
in the SED fitting. In near-IR, we used the fluxes obtained in 2012 
to include both {\it H} and {\it K} band fluxes.
Since YSO1 and YSO2 are not resolved at longer ($>$22~\micron) wavebands,
we first determined the relative contributions to the total fluxes from each YSO
by adopting the fraction factors of YSO1, $x$ (for WISE 22~\micron) and 
$y$ (for PACS 70~\micron), defined in this way: 
when the fraction factor is 1, a hundred per cent of the flux at 
the corresponding waveband comes from YSO1; when the fraction factor 
is 0, zero per cent of the flux at the corresponding waveband comes from 
YSO1, i.e., all flux comes from YSO2.
Changing $x$ and $y$ in a range between 0 and 1 with an interval 
of 0.05, we simultaneously fitted the SEDs of the two YSOs 
with a fixed distance to find the model sets with total reduced 
chi-squares $\lesssim 3$.
Figure~\ref{fig6} shows the reduced chi-square contours of
the fitting with the distance of 1.7~kpc, the distance to IRDC G53.2 \citep{kim15}; 
from these contours, we have found 
the best fraction factors of $x=0.775 \pm 0.08$ and $y=0.85 \pm 0.1$.
Fittings with smaller/larger distances between 1.5 and 2.0~kpc also gave
the similar fraction factors of $x \sim 0.8$ and $y \sim 0.8$ often with 
larger chi-squares, so we have adopted the fraction factors obtained
from the $d$=1.7~kpc models.
Applying these fraction factors, $x$=0.775 and $y$=0.85, to 
the 22~\micron\ and 70~\micron\ fluxes (e.g., $f_{\rm 22,YSO1} = x f_{\rm 22,total}$, 
$f_{\rm 22,YSO2} = (1-x) f_{\rm 22,total}$), 
we fitted the SED of each YSO again to find the best SED models.
The IRAM 1.2~mm flux \citep[the integrated 1.2~mm flux from][]{rath06}
was used as a upper limit after the fraction factor $y$ (the same factor 
as PACS 70~\micron) was applied.
In the fitting, 
considering the uncertainty in background variations and the fraction factors,
we assumed the flux uncertainty of 10\% that is 
larger than photometric errors.
A free parameter of external extinction $A_{\rm V}$ was allowed to
vary between 0 and 100~mag as the extinction toward \core\ is 
expected to be large (Section~\ref{sec:lum}). 
We used the extinction model of \citet{kim94} included in the SED Fitter,
a model that fitted a typical Galactic interstellar medium curve modified for 
the mid-IR extinction properties derived by \citet{indebetouw05} 
\citep{robitaille07}.
Although distance can be also given as a free parameter,
we fixed the distance as 1.7~kpc to reduce the number of free 
parameters because this distance was independently derived from
$^{13}$CO data \citep{kim15} and $\sim$10\% of uncertainty
in distance does not significantly affect the fitting results 
(see below).


\begin{deluxetable}{lcc}
\tabletypesize{\scriptsize}
\tablecaption{
Mid- and Far-IR Fluxes of 
YSO1 and YSO2 (in mJy) \label{tbl4}}
\tablewidth{0pt}
\tablehead{
\colhead{Data} & \colhead{YSO1} & \colhead{YSO2}}
\startdata
IRAC [3.6~\micron] & 435.1 $\pm$ 42.5  & 78.8 $\pm$ 14.8  \\
IRAC [4.5~\micron] & 3063.0 $\pm$ 310.3 &  182.0 $\pm$ 27.2 \\
IRAC [5.8~\micron] & 4425.0 $\pm$ 203.8 &  384.3 $\pm$ 16.3 \\
IRAC [8.0~\micron] & \nodata  &  656.2 $\pm$ 26.6 \\
WISE [22~\micron] & \multicolumn{2}{c} {37390. $\pm$ 206.6} \\
PACS [70~\micron] & \multicolumn{2}{c} {307481. $\pm$ 10.9}  \\
IRAM [1.2~mm] & \multicolumn{2}{c} {1770.}  \\
\enddata
\tablecomments{YSO1 and YSO2 are not resolved at $>$22~\micron. 
The PACS 70~\micron\ flux and IRAC 1.2~mm flux are from \citet{traficante15} and
\citet{rath06}, respectively.}
\end{deluxetable}


\begin{figure}
\includegraphics[width=1\textwidth]{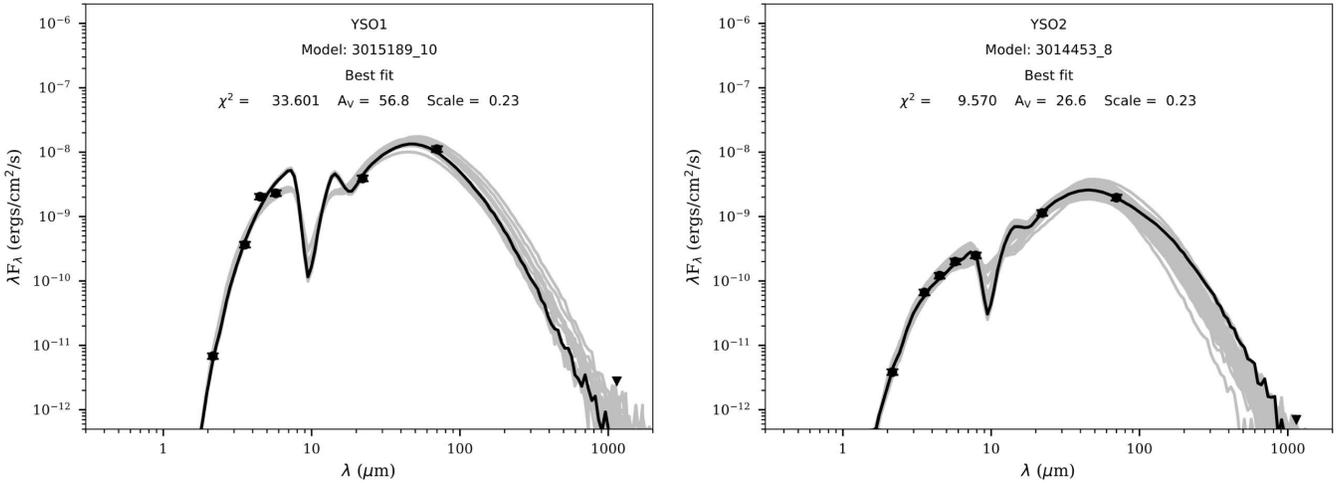}
\caption{SEDs of YSO1 and YSO2 with the fitted SED models.
Black lines present the best fitted model, and gray lines 
are the models satisfying the criterion of the goodness-of-fit, 
$\chi^2 - \chi^2_{\rm best} < 3 \times n_{\rm data}$. Scale of 0.23
is the distance of 1.7~kpc in log scale.
\label{fig7}}
\end{figure}

The SED fitting results are shown in 
Figure~\ref{fig7}. The black lines present the best-fitting models, and
the gray lines present ``good" models satisfying the criterion of 
$\chi^2 - \chi^2_{\rm best} < 3 \times n_{\rm data}$, where
$\chi^2$ is total chi-square from fitting, $\chi^2_{\rm best}$
is the total chi-square of the best-fitting model, and $n_{\rm data}$
is the number of data points used in fitting.
The fitted parameters of the best models and the parameter ranges
of good models are listed in Table~\ref{tbl5}.
The evolutionary stages in the table have been determined by
the Stage classification scheme of \citet{robitaille06}, 
a scheme that is defined from the ratio of envelope accretion
rate ($\dot{M}_{\rm env}$) or disk mass ($M_{\rm disk}$) 
to central source mass ($M_{\star}$):
Stage I (including Stage 0) for the ones
with $\dot{M}_{\rm env}/M_{\star} > 10^{-6}~{\rm yr^{-1}}$; 
Stage II for the ones 
with $\dot{M}_{\rm env}/M_{\star} < 10^{-6}~{\rm yr^{-1}}$ 
and $M_{\rm disk}/M_{\star} > 10^{-6}$;
and Stage III for the ones
with $\dot{M}_{\rm env}/M_{\star} < 10^{-6}~{\rm yr^{-1}}$ 
and $M_{\rm disk}/M_{\star} < 10^{-6}$.
The criterion of $\chi^2 - \chi^2_{\rm best} < 3 \times n_{\rm data}$, we 
used to select good models, is the same as the one
defined in \citet{robitaille07}. Although this criterion is 
arbitrary and fairly loose in statistical aspects, as \citet{robitaille07} 
pointed out, it provides a range of acceptable fits to the eye and 
reasonable constraints. Considering the sparse coverage of 14-dimensional 
parameter space, the uncertainties of the models, and other 
realistic factors such as intrinsic variability or asymmetrical geometry
of YSOs, this criterion also would prevent the risk of overinterpretation 
of SEDs from using a stricter criterion \citep{robitaille07, forbrich10}.


\begin{deluxetable}{lccccccc}
\tabletypesize{\scriptsize}
\tablecaption{SED Fitting Parameters of YSO1 and YSO2 from Good Models
selected by
$\chi^2 - \chi^2_{\rm best} < 3 \times n_{\rm data}$ \label{tbl5}}
\tablewidth{0pt}
\tablehead{
\colhead{ } &
\multicolumn{3}{c}{YSO1} & \colhead{} &
\multicolumn{3}{c}{YSO2} \\
\cline{2-4} \cline{6-8} \\
\colhead{Parameters} & \colhead{min} & \colhead{best} & \colhead{max} &
\colhead{ } & \colhead{min} & \colhead{best} & \colhead{max}}
\startdata
Central source mass (\msol) & 
7.94 & 9.96 & 10.1 & & 5.27 & 5.39 & 7.64 \\
Central source age (yr) & 
$1.18 \times 10^{3}$ & $3.21 \times 10^{3}$ & $2.53 \times 10^{5}$ & &
$2.62 \times 10^{3}$ & $2.62 \times 10^{3}$ & $9.79 \times 10^{5}$ \\
Total luminosity ($L_{\sun}$) &
$1.87 \times 10^{3}$ & $2.20 \times 10^{3}$ & $4.42 \times 10^{3}$ & &
$2.87 \times 10^{2}$ & $4.60 \times 10^{2}$ & $1.46 \times 10^{3}$ \\
Central source temperature (K) &
$4.13 \times 10^{3}$ & $4.32 \times 10^{3}$ & $2.46 \times 10^{4}$ & &
$4.17 \times 10^{3}$ & $4.17 \times 10^{3}$ & $2.01 \times 10^{4}$ \\
Envelope accretion rate ($M_{\sun}~{\rm yr^{-1}}$) &
$2.88 \times 10^{-5}$ & $1.90 \times 10^{-4}$ & $2.63 \times 10^{-4}$ & &
$6.29 \times 10^{-8}$ & $6.77 \times 10^{-5}$ & $1.69 \times 10^{-4}$ \\
Disk mass (\msol) &
0 & 0 & $5.68 \times 10^{-2}$ & &
$9.03 \times 10^{-5}$ & $1.05 \times 10^{-2}$ & $4.78 \times 10^{-1}$ \\
Interstellar extinction, $A_{\rm V}$ (mag)\tablenotemark{a} & 
40.42 & 56.81 & 60.96 & & 15.67 & 26.59 & 54.58 \\
Stage & & I & & & & I/II & 
\enddata
\tablenotetext{a}{This $A_{\rm V}$ only accounts for external foreground 
extinction and does not include the self-extinction 
by circumstellar dust.}
\end{deluxetable}


By the criterion of $\chi^2 - \chi^2_{\rm best} < 3 \times n_{\rm data}$,
17 and 37 good models have been selected for YSO1 and YSO2, respectively,
and all of the good models fairly well explain the observed SEDs of the two YSOs 
as seen in Figure~\ref{fig7}, with the reduced chi-squares of
4.8--7.1 (YSO1) and 1.2--4.1 (YSO2). The parameter ranges
presented in Table~\ref{tbl5} are mostly within one or two order of magnitude
except the envelope accretion rate and disk mass of YSO2, giving an ambiguous
evolutionary stage between Stage I and II. 
These large parameter ranges can be improved if we have more data points,
particularly far-IR/submillimeter data. Fluxes at longer wavebands 
significantly affect the determination of envelope accretion rate and disk mass 
as pointed out in \citet{robitaille07}.
We also tried fitting with distance varying in a range
between 1.5 and 2~kpc. The increased number of free parameters
increased the number of good models to 40 for YSO1 and 81 for YSO2, 
but their parameter ranges agree well with the ranges 
in Table~\ref{tbl5}, confirming that the uncertainty of distance 
insignificantly affects the fitting results. 
The mean distances derived from the fitting are 1.65~kpc and 1.71~kpc 
for YSO1 and YSO2, respectively, so the distance of 1.7~kpc we assumed 
is reasonable.

As indicated in Table~\ref{tbl5}, the young age and high envelope accretion 
rate of YSO1 confirm that it is a high-mass protostar, as previously implied
by the detection of 6.7~GHz class II methanol 
maser \citep{pandian11}; YSO2, on the other hand, is rather
close to an intermediate-mass YSO with lower mass.
The evolutionary stage of Stage I, consistent with the class 
determined from the spectral index, indicates that either YSO1
or YSO2 can drive the outflow. 
Some models of YSO2 fall in Stage II with lower
envelope accretion rate, but 80\% of the models correspond to
Stage I.

We note a large difference of interstellar
extinction ($A_{\rm V}$) between YSO1 and YSO2. As this parameter
$A_{\rm V}$ only accounts for external foreground extinction excluding
the self-extinction by circumstellar dust, the two YSOs at the same
distance are generally expected to have similar $A_{\rm V}$.
But in Table~\ref{tbl5}, $A_{\rm V}$ of YSO1 is about two times 
larger than $A_{\rm V}$ of YSO2, although the maximum $A_{\rm V}$
is comparable.
This is likely because YSO1 is almost at the center of the core
where the extinction value is the maximum and
YSO2 is a little away from the center where
the extinction value is expected to be smaller.
For example, 
the radial profiles of the mass surface density of 
IRDC cores at the distance of 2--3~kpc derived by mid-IR
extinction technique \citep{butler12} show that the mass surface densities
are the maximum at $r < 1\arcsec$ and gradually decrease to
$\sim$40--60\% of the maximum values at $r \sim 10\arcsec$
\citep[Figures 5--12 of][]{butler12}.
If \core\ has a similar mass surface density profile to those IRDC cores,
the mass surface density would decrease by a half
at the position of YSO2 separated by $\sim$8$\arcsec$, and therefore, 
the difference of $A_{\rm V}$ between YSO1 and YSO2
from the SED fitting is acceptable.
Additionally, we compare the extinction of YSO1 and YSO2 derived
by their near-IR color. The $A_{\rm V}$ obtained by applying 
the {\it H}- and {\it K}-band magnitudes in Table~\ref{tbl2} to 
the Equation~1 of \citet{cooper13} is
$\sim$100 and $\sim$60~mag for YSO1 and YSO2, respectively. 
Since extinction from near-IR color includes dust-excess from circumstellar 
material (self-extinction), the derived $A_{\rm V}$ of the two YSOs are larger 
than $A_{\rm V}$ from the SED fitting or $N(^{13}{\rm CO})$, 
but they show a difference by about a factor of two, consistent with 
the $A_{\rm V}$ difference found in the SED fitting results.

\section{Origin of the \core\ Outflow}\label{sec:origin}

\begin{figure}
\includegraphics[width=1\textwidth]{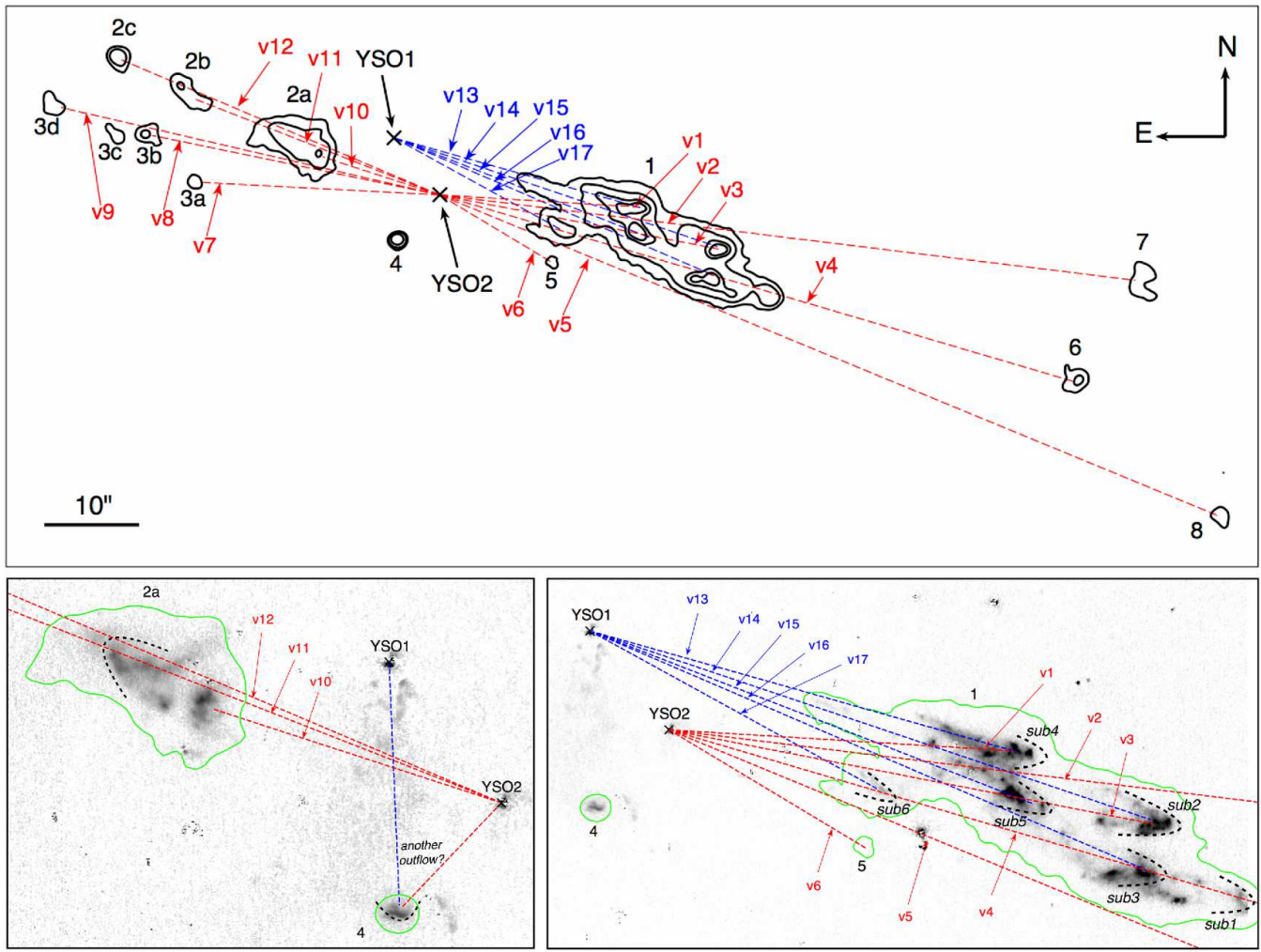}
\caption{Top: \htwo\ outflow contours ($3\sigma, 10\sigma, 45\sigma$, and 
$80\sigma$ above the background in Figure~\ref{fig2}) and the vectors
that trace the outflow emission features by assuming the driving source
as YSO1 (blue; from v13 to v17) and YSO2 (red; from v1 to v12).
Bottom: Continuum-subtracted Subaru/IRCS \htwo\ images around the
flow \no1 and flow \no2a with the outflow vectors. 
Possible bow-shock tips are drawn by black dashed-lines, and 
sub-flows in the flow \no1 are labeled from {\it sub1} to {\it sub6}.
\label{fig8}}
\end{figure}

The \core\ outflow is likely associated with the YSOs
at the outflow center.
Physical properties of the central YSOs examined in the previous 
section indicate that the both are capable of ejecting outflows,
but which one is indeed driving the outflow is unclear. 
As described in Section~\ref{sec:apparent},
the overall morphology of the \core\ outflow is bipolar with one
lobe much brighter than the other. This bipolar morphology is in general
interpreted as the outflow ejected from a single source 
with the brighter lobe blueshifted and the fainter lobe redshifted.
If the whole outflow emission only originates from a single source,
YSO2 seems to better explain the outflow morphology than YSO1.
In Figure~\ref{fig8}, we present vectors tracing
the \htwo\ features on the \htwo\ emission contours at
the top panel and on the continuum-subtracted Subaru/IRCS images 
at the bottom panels that show a detailed structure of the flows \no1 and
\no2a.
As drawn by the vectors from v1 to v12 in red color, YSO2 fairly
well explains all of the emission features except \no4 as 
the outflow with PA $\sim74\arcdeg$ and opening 
angle $\sim 30 \arcdeg$.
The red vectors present at least three and five flows with
different directions to the northeast and to the southwest, respectively. 
In particular, the flow \no1 is clumpy and consists of several bow-shock flows 
as seen in the bottom right panel of Figure~\ref{fig8}. Similar structures of
multiple bow shocks have been observed in the high-resolution optical 
and near-IR images of HH1/HH2 and explained by thermal instabilities 
from the shock front running into inhomogeneous and perhaps rather 
dense ambient gas or by variability in jet direction \citep{hester98,davis00}.
These radially propagating flows with bow-shock tips in the flow \no1 are in part
similar to the ``\htwo\ fingers" of the Orion BN/KL outflow \citep{bally15} produced
by an explosive outflow with simultaneously ejected multiple jets from a high-mass YSO;
however, when we consider the high degree of collimation and the small opening angle 
compared to the BN/KL outflow, a more feasible interpretation is
multiple precessing jets. For example,
if the outflow ejected from YSO2 experienced precession, 
the observed outflow morphology can be explained by at least two precessing jets.

Assuming YSO2 as a driving source,
we can simply explain the entire outflow as discussed above, 
but it only describes the geometrical morphology that is projected 
on the sky, so we cannot rule out the possible contribution from YSO1
to the outflow. 
As the blue vectors from v13 to v17 in Figure~\ref{fig8} show,
YSO1 well explains the \htwo\ emission features in the southwest. 
(The vectors tracing the features from \no5 to \no8 are
not drawn for simplification.)
In the bottom right panel of the figure,
some sub-flows in the flow \no1 are even better explained by YSO1, 
for example a faint bow-shock feature {\it sub6} traced by v17, or 
a jet-like feature along v13.
In this case, the outflow is defined by PA $\sim67\arcdeg$ and 
opening angle $\sim 27 \arcdeg$.
Therefore, it can be suggested that the outflow toward southwest at least
in part originates from YSO1 while its counter jet that is likely redshifted 
is not observed due to larger extinction to the opposite side.
The \htwo\ emission in the northeast, on the other hand,
are hardly traced by a vector from YSO1. If they originated from
YSO1, a jet would have been ejected toward southeast and
bent by $\sim$90$\arcdeg$ toward northeast.
This requires the outflow to have experienced a high degree
of precession, but it is not likely because curved or wiggly structures
expected from precession are not seen among the other features.
Another possible explanation for the NE flow in a relation with YSO1
is that the outflow axis has an inclination in the way that the NE axis
is toward us, i.e., the NE lobe is blueshifted and the SW lobe
is redshifted.  
This possibility conflicts with the general expectation that 
the blueshifted lobe is brighter than the redshifted one because of 
lower column density along the line of sight, but such expectation may 
not be applied if there is a region with locally enhanced extinction.
The NE flow is closer to YSO1, i.e., the center of the core, 
than the SW flow, so 
the NE side is expected to have larger extinction than the SW side 
since extinction increases toward the center of the core as discussed 
in Section~\ref{sec:sed}.
Therefore, the NE flow can be a blueshifted lobe that is fainter 
than the other side due to locally larger extinction.

On the other hand, 
we can think of a possibility that there is another outflow-driving source
in the core besides YSO1 and YSO2 that has not yet been detected.
As a protostar can eject outflows from very young evolutionary phase 
surrounded by a thick envelope, outflow-driving YSOs are often 
so deeply embedded that they are not observed in near- or mid-IR but
only observed in (sub)millimeter \citep[e.g., LkH$\alpha$ 234 region;][]{fuente01}. 
In addition, recent ALMA observations have revealed that a massive core
in fact is composed of several lower-mass cores embedded in a dust filament,
cores that can be only resolved at high angular resolution 
of $\lesssim 1\arcsec$ \citep[e.g., G35.20-0.74;][]{sanchez14}, suggesting
that \core\ also possibly contains undetected, deeply embedded 
protostars driving the \htwo\ outflow.
We note that the outflow luminosity derived in Section~\ref{sec:lum}
implies $\sim 10^4 < L_{\rm bol}/L_{\sun} < 10^6$ for the driving source
on the relation of $L_{\rm H_2} \propto L_{\rm bol}^{0.6}$ \citep{caratti15}.
If this empirical relation works here,
the luminosity of YSO1 ($\sim 2 \times 10^{3}~L_{\sun}$) 
and YSO2 ($\sim 0.5 \times 10^{3}~L_{\sun}$) inferred from SED fitting 
does not seem to be enough to explain the observed outflow luminosity.
This may imply the possible presence of another outflow-driving source 
that may be massive enough to solely eject the observed \htwo\ outflow, 
but it is more feasible that the \core\ outflow is composed of multiple outflows 
of different origins including YSO1 and YSO2, for they are all very young YSOs 
in early phase expected to eject outflows.

Lastly, we discuss about the \htwo\ emission feature \no4.
In the UKIRT image (Figure~\ref{fig2}), \no4 appears as a compact knot, 
but in the Subaru image (the bottom panels of Figure~\ref{fig8}), it appears 
as a small, thin filament with a curvature similar to a bow-shock tip
of which apex is well connected to either YSO1 or YSO2.
As discussed in Section~\ref{sec:pa},
the PA of \no4 with respect to either YSO is very different from
the PAs of the other emission features or the overall PA of the outflow,
suggesting that there may be another outflow
differentiated from the NW-SE outflow.
There are also a small, elongated feature at the west of \no4 
at the level of one sigma above the background (Figure~\ref{fig2}) and
faint features between YSO1 and \no4 (Figure~\ref{fig8}) 
although the latter are not clear whether they are real \htwo\ emission
or residual nebula emission left from continuum subtraction.
If the faint elongated feature is also a part of another outflow with \no4, 
the outflow direction is from north to south and the driving source 
is likely YSO1. 

\section{Summary and Conclusion}\label{sec:summary}

We have presented a parsec-scale \htwo\ outflow discovered 
in the IRDC core \core\ at the distance of 1.7~kpc. 
The overall morphology of the outflow is bipolar along the NE-SW direction in 
the \htwo\ 1-0\,S(1) 2.12~\micron\ image.
At the outflow center, there are two Class I YSOs (YSO1 and YSO2) separated
by $\sim$8$\arcsec$; we consider both to be putative outflow-driving sources  
based on their IR colors and location.
We derived the physical parameters of the \htwo\ outflow and the central
YSOs using the \htwo\ images and the broad-band near-IR images, 
and have discussed their association. 
Our results and the implications on the properties and origin of the outflow
can be summarized as follows. 

\begin{enumerate}
\item The outflow is bipolar from northeast to southwest with the SW flow is 
much brighter than the NE flow, but the detailed structure is
composed of several discrete flows and knots. From the UKIRT \htwo\ image,
we identified 13 emission features using the threshold of three sigma above 
the background. 
The outflow, with the total length of $\sim$130$\arcsec$ or $\sim$1~pc at 1.7~kpc, 
is relatively long compared with the observed protostellar outflows from 
low-mass YSOs. The dynamical age, although it highly depends on outflow 
velocity, is from several thousand to a few tens of thousand years.
Some of the \htwo\ emission features are well aligned and show time gaps 
about 1,000~yr at the outflow velocity of 80~\kms; a few thousand years of
time gaps are comparable to the time gaps reported 
in previous studies \citep[e.g,][]{ioannidis12b,froebrich16} and suggest episodic 
or non-steady mass ejection history. The position angle of the outflow
is uncertain without a confirmed driving source but is around 70$\arcdeg$.

\item The total extinction-corrected \htwo\ luminosity of the outflow 
is $L_{\rm H_{2}} \sim $~(6--150)~$L_{\sun}$.
We adopt an Av of between 15 and 50 mag, based on the average
optical depth of a larger scale {\it Spitzer} dark cloud including \core\ 
and $^{13}$CO column density, respectively.
If the whole outflow is ejected from a single source, the observed \htwo\ luminosity 
that is several times larger than the luminosity of the outflows from 
low-/intermediate-mass YSOs \citep{caratti06,ioannidis12b} 
implies a high-mass outflow-driving source for the \core\ outflow. 
The empirical relationship between the \htwo\ luminosity of the outflow and 
the bolometric luminosity of the driving source 
\citep[$L_{\rm H_2} \propto L_{\rm bol}^{\alpha}$ with $\alpha \sim 0.6$;][]{caratti15} 
also suggests that the driving source of the \core\ outflow is massive with 
$\sim 10^4 < L_{\rm bol}/L_{\sun} < 10^6$.

\item We identified compact, faint \feii\ emission features from the high-resolution 
Subaru/IRCS image. The \feii\ emission is marginally detected inside 
the \htwo\ flows with the area of $\sim$1~arcsec$^{2}$ and 
the surface brightness about ten times smaller than the \htwo\ brightness.
But it is difficult to conclude that the detected \feii\ emission is associated 
with the \htwo\ outflow because \feii\ knots are generally observed 
at the tips of the \htwo\ jets rather than behind the \htwo\ bow shocks.
Since the \htwo\ and \feii\ lines arise from different shock origins,
the marginal detection of \feii\ emission may indicate that slow, C-type 
shocks are dominant in the \core\ outflow, although deeper imaging 
observations with higher sensitivity or spectroscopic observations
are required to derive the physical conditions of the region and confirm
shock properties.

\item Both central YSOs show photometric variability in {\it H}- and {\it K}-bands 
between several years. The available data are limited to present the full variability, 
but high variability amplitudes of $\gtrsim$1~mag suggest that they can be
eruptive variable YSOs with episodic outbursts.
The SED fitting of the two YSOs shows that 
both YSOs are indeed in the early evolutionary stage with high envelope accretion 
rates of $10^{-5}$--$10^{-4} M_{\sun}~{\rm yr}^{-1}$, implying that the both are
proper candidates of the outflow-driving source.
The masses inferred from the best SED fitting models are $\sim$10~\msol\ 
and $\sim$5~\msol\ for YSO1 and YSO2, respectively. 
This supports the association between the \htwo\ outflow
and a high-mass YSO, and also confirms high-mass star formation occurring
in the IRDC core.

\item The \core\ outflow is most likely associated with the two central YSOs.
The young evolutionary stages of both YSOs support their association,
but which one is driving the outflow is still unclear. 
YSO2 well explains the geometrical morphology of the outflow
as a single-source origin. But we cannot rule out the possible contribution 
from YSO1 because it also well describes the outflow emission in the southwest, 
and may explains the emission in the northeast as well if the NE axis of the outflow
is toward us. The outflow, by assuming either YSO as a driving source,
can be defined by PA $\sim 70\arcdeg$ and opening angle $\sim 30\arcdeg$;
the radial flows of different directions with bow-shock tips may suggest 
multiple precessing jets.
In addition, we consider a possibility of the presence of another outflow-driving 
source very deeply embedded in the core that has not been detected
in near- and mid-IR but could be detected in the submillimiter with high-spatial resolution.

\item Our results show that the \core\ outflow has a complicated 
morphology with more than one outflow-driving source candidates.
One of the \htwo\ features with very different PA from the other
features even raises a possibility that there is another outflow, implying 
that the \core\ outflow is a combination 
of multiple outflows of several different origins.
Our study also implies that   
the parsec-scale, collimated \htwo\ outflow, 
at least in part, originates from a massive ($\sim$10~\msol) YSO 
as well as an intermediate-mass ($\gtrsim$5~\msol) YSO, 
suggesting intermediate- to high-mass star formation
by mass accretion via disks as low-mass star formation.
Follow-up observations
particularly to obtain the kinematic information of the outflow and
to search for molecular outflows directly ejected from the central YSOs 
will be necessary in order to confirm these possibilities and fully understand
the outflow characteristics in future.

\end{enumerate}

\acknowledgments

We thank Lee, J.-J. who helped in obtaining the Gemini/NIRI data.
This work is based on observations made with the Spitzer Space Telescope, 
which is operated by the Jet Propulsion Laboratory, California Institute of 
Technology under a contract with NASA.
This work was supported by NRF(National Research Foundation of Korea) 
Grant funded by the Korean Government
(NRF-2012-Fostering Core Leaders of the Future Basic Science Program).
This work was supported by the National Research Foundation of Korea (NRF) 
grant funded by the Korea Government (MSIP) (No. 2012R1A4A1028713).
This work was supported by K-GMT Science Program (PID:GN-2015B-Q-16) 
of Korea Astronomy and Space Science Institute (KASI).

\facilities{UKIRT (WFCAM), Subaru (IRCS), Gemini:Gillett (NIRI)}
\software{IRAF v2.16 \citep{tody86,tody93}, STARFINDER \citep{diolaiti00}}



\end{document}